\documentclass[fleqn,usenatbib]{mnras}

\usepackage{natbib}
\usepackage{soul}

\usepackage[T1]{fontenc}
\usepackage{ae,aecompl}

\usepackage{outlines}
\usepackage{color}

\usepackage{graphicx}	% Including figure files
\usepackage{amsmath}	% Advanced maths commands
\usepackage{amssymb}	% Extra maths symbols
\usepackage{bm}	        % bold font in math
\usepackage{newtxtext,newtxmath}
\usepackage{hyperref}

\usepackage{ulem}
\usepackage{natbib}

\def\mrm{\mathrm}

\defcitealias{Zu2021}{Z21}
\defcitealias{Zibetti2005}{Z05}

\newcommand{\rsoi}{R_{\mathrm{SOI}}}
\newcommand{\sigbi}{\Sigma_*^{\texttt{B+I}}}
\newcommand{\sigm}{\Sigma_m}
\newcommand{\sigg}{\Sigma_g}
\newcommand{\sigdev}{\Sigma_*^{\texttt{deV}}}
\newcommand{\sigtr}{\Sigma_*^{\texttt{tran}}}

\newcommand{\mubi}{\mu^{\texttt{B+I}}}

\newcommand{\sbmag}{\mathrm{mag}/\mathrm{arcsec}^{2}}
\newcommand{\sqarcsec}{\mathrm{arcsec}^{2}}

\newcommand{\avg}[1]{\left\langle #1 \right\rangle}

\newcommand{\dd}{\mathrm{d}}

\newcommand{\msbcg}{M_*^{\texttt{BCG}}}
\newcommand{\mtwenty}{M_{*,20\kpc}}

\newcommand{\mh}{M_h}

\newcommand{\mpc}{\mathrm{Mpc}}
\newcommand{\hmpc}{h^{-1}\mathrm{Mpc}}

\newcommand{\kpc}{\mathrm{kpc}}
\newcommand{\msol}{M_{\odot}}
\newcommand{\hhmsol}{h^{-2}M_{\odot}}
\newcommand{\ds}{\Delta\Sigma}

\newcommand{\pcen}{p_{\mathrm{cen}}}
\newcommand{\pmem}{p_{\mathrm{mem}}}

\newcommand\redmapper{\texttt{redMaPPer}}

\newcommand{\rom}[1]{\uppercase\expandafter{\romannumeral #1\relax}}
\title[Decomposition of the Diffuse Cluster Light]{The Sphere of
Influence of the Bright Central Galaxies in the Diffuse Light of SDSS
Clusters}
\author[Chen 2021]
{Xiaokai Chen$^{1}$,
Ying  Zu$^{1, 2}$\thanks{E-mail: yingzu@sjtu.edu.cn},
Zhiwei Shao$^{1}$,
Huanyuan Shan$^{3, 4}$
\\ \\
$^{1}$Department of Astronomy, School of Physics and Astronomy, Shanghai Jiao Tong
University, Shanghai 200240, China\\
$^{2}$Shanghai Key Laboratory for Particle Physics and Cosmology, Shanghai Jiao Tong
University, Shanghai 200240, China\\
$^{3}$Shanghai Astronomical Observatory (SHAO), Nandan Road 80, Shanghai 200030, China \\
$^{4}$University of Chinese Academy of Sciences, Beijing 100049, China
}

\date{Accepted XXX. Received YYY; in original form ZZZ}

\pubyear{2021}

\begin{document}

\label{firstpage}
\pagerange{\pageref{firstpage}--\pageref{lastpage}}
\maketitle

% Abstract of the paper
\begin{abstract}
    The bright central galaxies~(BCGs) dominate the inner portion of the
    diffuse cluster light, but it is still unclear where the intracluster
    light~(ICL) takes over. To investigate the BCG-ICL transition, we stack
    the images of ${\sim}3000$ clusters between $0.2{<}z{<}0.3$ in the SDSS
    $gri$ bands, and measure their BCG+ICL stellar surface mass profile
    $\Sigma_{*}^{\texttt{B+I}}$ down to
    $3{\times}10^4\,M_{\odot}/\mathrm{kpc}^{2}$ at
    $R{\simeq}1\,\mathrm{Mpc}$~(${\sim}32$ mag/arcsec$^2$ in the $r$-band).
    We develop a physically-motivated method to decompose
    $\Sigma_{*}^{\texttt{B+I}}$ into three components, including an inner de
    Vaucouleurs' profile, an outer ICL that follows the dark matter
    distribution measured from weak lensing, and an intriguing transitional
    component between 70 and 200 kpc. To investigate the origin of this
    transition, we split the clusters into two subsamples by their BCG
    stellar mass $M_*^{\mathrm{BCG}}$~(mass enclosed roughly within 50 kpc)
    while making sure they have the same distribution of satellite
    richness. The $\Sigma_{*}^{\texttt{B+I}}$ profiles of the two subsamples
    differ by more than a factor of two at R < 50 kpc, consistent with
    their 0.34 dex difference in $M_*^{\mathrm{BCG}}$, whereas on scales
    beyond 400 kpc the two profiles converge to the same amplitudes,
    suggesting a satellite-stripping origin of the outer ICL.  Remarkably,
    however, the discrepancy between the two $\Sigma_{*}^{\texttt{B+I}}$
    profiles persist at above $50\%$ level on all scales below 200 kpc, thereby revealing
    the BCG sphere of influence with radius $R_{\mathrm{SOI}}{\simeq}$ 200
    kpc. Finally, we speculate that the surprisingly large sphere of
    influence of the BCG is tied to the elevated escape velocity profile
    within $r_s$, the characteristic radius of the dark matter haloes.
\end{abstract}
% Select between one and six entries from the list of approved keywords.  Don't make up
% new ones.

%%%%%%%%%%%%%%%%%%%%%%%%%%%%%%%%%%%%%%%%%%%%%%%%%%

%%%%%%%%%%%%%%%%% BODY OF PAPER %%%%%%%%%%%%%%%%%%

%%%%%%%%%%%%%%%%%%%%%%%%%%%%%%%%%%%%%%%%%%%%%%%%%%

\vspace{1in}

\section{Introduction}
\label{sec:intro}

The intracluster light~(ICL) is primarily produced by stray stars that have
escaped individual galaxies but remain gravitationally bound to the cluster
potential~\citep{Zwicky1937, deVaucouleurs1970, Welch1971, Melnick1977,
Feldmeier2004, Mihos2005, Krick2007}. As the by-product of galaxy
interactions within clusters~\citep[for a recent review;
see][]{Contini2021a}, these free-floating stars are key to unlocking the
assembly history of the bright central galaxies~(BCGs\footnote{We do not
conform to the more commonly-adopted nomenclature of ``brightest cluster
galaxies'', because the BCGs in this work are likely the
central galaxies but not necessarily the brightest.}) with
future deep imaging surveys like LSST~\citep[{\it Rubin};][]{Ivezic2019},
{\it Euclid}~\citep{Laureijs2011}, {\it Roman}~\citep{Spergel2015}, and the
Chinese Survey Space Telescope~\citep[{\it CSST};][]{Gong2019}. However, it
is unclear whether there exists an ICL component that is physically
distinct from the diffuse stellar envelope of the BCG~\citep{Kormendy1974},
and if so, where the ICL begins and the sphere of influence of the BCG
ends~\citep{Doherty2009}. In this paper, we examine the BCG+ICL stellar
surface mass profile $\sigbi$ of a large sample of clusters observed by the
Sloan Digital Sky Survey~\citep[SDSS;][]{York2000}, in hopes of finding a
more physically-motivated method of decomposing the $\sigbi$ profile and
the BCG ``sphere of influence'' within the diffuse cluster light.

In hydrodynamical simulations, methods for kinematically decomposing
BCG vs. ICL have been developed based on the apparent bimodal
distribution of the intracluster stellar velocities~\citep{Dolag2010,
Cui2014}. However, such exquisite methods developed for simulations are not
applicable in the observations of distant clusters where stellar velocities
are inaccessible~\citep[but see][for kinematic studies of ICL in local
clusters]{Arnaboldi1996, Romanowsky2012, Longobardi2015, Spiniello2018,
Gu2020}. As a result, traditional methods of decomposition usually assume
that the diffuse light below some arbitrary surface brightness~(SB) limit
belongs to the ICL~\citep{Rudick2011, Burke2012, Presotto2014, Tang2018,
Furnell2021}, or describe the BCG+ICL SB profile $\mubi$ as the sum of
multiple S\'ersic components~\citep{Gonzalez2005, Seigar2007, Donzelli2011,
Cooper2015, Zhang2019, Montes2021, Kluge2021}. Despite the incoming deep
photometric dataset from LSST+{\it Euclid}+{\it Roman}+{\it CSST}, a
well-defined physical decomposition of the BCG+ICL profile is still
lacking.

Recently, several studies proposed that the ICL in the outer region of
clusters should follow the distribution of dark matter, due to the fact
that the stray stars and dark matter particles are both collisionless
tracers of the cluster potential~\citep{Montes2019, AlonsoAsensio2020,
Contini2020, SampaioSantos2021}. By the same token, the outer ICL may also follow the
distribution of satellite galaxies, whose mass loss contributes
significantly to the ICL~\citep{Purcell2007, Martel2012, Contini2014,
Morishita2017}. In this paper, we not only measure the BCG+ICL stellar
surface mass density profile $\sigbi$ from image stacking, but also the
total surface mass density profile $\sigm$ from cluster weak lensing, as
well as the galaxy surface number density profile $\sigg$ from
cluster-galaxy cross-correlation functions. By comparing the $\sigbi$
profile with $\sigm$ and $\sigg$, we develop a physically-motivated method
for the decomposition of the diffuse cluster light without resorting to
arbitrary SB limits or multiple S\'ersic profiles.

In theory, since the intracluster stars are unlikely born {\it in
situ}~\citep{Puchwein2010, Melnick2012}, the build-up of the diffuse light
should be closely linked to the dynamical evolution of the BCG and
satellite galaxies, including: (1) mergers of satellites with the
BCG~\citep{Murante2007, Burke2015}, (2) tidal disruption/stripping of
satellites in the central region of the cluster~\citep{Purcell2007,
Wetzel2010, Montes2014, DeMaio2015, Contini2018}, (3) tidal and
ram-pressure stripping of infalling satellites~\citep{Rudick2009,
Contini2014, Montes2018, DeMaio2018, Contini2019, JimenezTeja2018}, and (4)
pre-processing within infalling groups~\citep{Willman2004,
SommerLarsen2006, Rudick2006}.  Among these processes, (1) and (2) can
simultaneously grow the BCG {\it and} the ICL within a Hubble time, while
(3) and (4) deposit stars only into the ICL without growing the BCG.  In
particular, the amount of ICL growth through (1) and (2) should be strongly
correlated with the observed BCG stellar mass $\msbcg$, but the ICL growth
induced by (3) and (4) would be tied instead to the number of satellite
galaxies within the cluster~(a.k.a., cluster richness $\lambda$).

Therefore, any physically meaningful decomposition should reveal at least
two distinct components of the diffuse cluster light, one {\it BCG-induced}
and the other {\it richness-induced}.  Obviously, the radial extent of the
BCG-induced portion can be regarded as the radius of the BCG sphere of
influence $\rsoi$ within the diffuse cluster light. In this paper, we
divide clusters into high and low-$\msbcg$ subsamples at fixed
$\lambda$~\citep[see also][hereafter referred to as
\citetalias{Zu2021}]{Zu2021}, and infer $\rsoi$ by comparing the $\sigbi$
profiles between the two subsamples. Using weak lensing, Z21 found that the
average halo concentration of the high-$\msbcg$ clusters is $\sim$10\% higher than
that of the low-$\msbcg$ clusters, but their average halo masses are the
same. Since the two populations have the same $\lambda$~(and average halo
mass) but differ only in $\msbcg$, we expect their $\sigbi(R)$ profiles to
have equal richness-induced contributions beyond $\rsoi$, but exhibit
distinct levels of BCG-induced diffuse mass below $\rsoi$ --- a potentially
smoking-gun detection of $\rsoi$.

This paper is organised as follows. We provide an overview of the cluster
catalogue and photometric images in \S\ref{sec:data}, setting the stage for
our BCG+ICL SB profile measurement pipeline in~\S\ref{sec:sb}. We describe
the measurement and the decomposition of the BCG+ICL stellar surface
density profile $\sigbi$ in~\S\ref{sec:sigma}, and then present the
observational detection of the BCG sphere of influence in~\S\ref{sec:soi}.
We conclude by summarizing our findings and looking to the future
in~\S\ref{sec:conc}.  Throughout the paper, we assume the
\citet{PlanckCollaboration2016} cosmology~($\Omega_m{=}0.31$,
$\sigma_8{=}0.816$, $h{=}0.677$), and convert all distances into physical
coordinates. We use $\lg x=\log_{10}x$ for the base-10 logarithm and $\ln
x=\log_{e}x$ for the natural logarithm.

\section{Data}
\label{sec:data}

\subsection{Cluster Catalogue}
\label{subsec:cls}

Following~\citetalias{Zu2021}, we employ the SDSS \redmapper~cluster
catalogue~\citep{Rykoff2014} derived by applying a red sequence-based
cluster finding algorithm to the SDSS DR8 imaging~\citep{Aihara2011}.
Briefly, \redmapper~iteratively self-trains a model of red-sequence
galaxies calibrated by an input spectroscopic galaxy sample, and then
attempts to grow a galaxy cluster centred about each photometric galaxy.
Once a galaxy cluster has been identified by the matched-filters, the
algorithm iteratively solves for a photometric redshift based on the
calibrated red-sequence model, and re-centres the clusters about the best
BCG candidates with the highest probability of being the central
galaxy~$\pcen$~\citep{Rykoff2014}.

For each detected cluster, \redmapper~applies an aperture of
${\sim}1\,\hmpc$~(with a weak dependence on richness $\lambda$), and assign
each galaxy within the aperture a membership probability $\pmem$. The
satellite richness $\lambda$ was computed by summing the $\pmem$ of all
member galaxy candidates, which roughly corresponds to the number of
red-sequence satellite galaxies brighter than $0.2\,L_*$. At
$\lambda{\geq}20$, the SDSS \redmapper~cluster catalogue is approximately
volume-complete up to $z{\simeq}0.33$, with cluster photometric redshift
uncertainties as small as $\delta(z){=}0.006/(1+z)$.  We select $4593$
clusters with $\lambda{\geq}20$ and redshifts between $0.2$ and
$0.3$~($\avg{z}{=}0.254$). The maximum redshift of $0.3$ is primarily set
by the requirement of sample completeness, and partly because the cosmic
dimming effect renders the detection of low-SB signals within SDSS
challenging at the higher redshift~($\mu \propto (1+z)^{-4}$).

We pick the galaxy with the highest $\pcen$ in each cluster as the BCG, and
derive an $i$-band cModel magnitude-based stellar mass $\msbcg$ for each
BCG.  In general, the SDSS model magnitudes are preferred for measuring the
color of extended objects like the BCGs~(hence a better mass-to-light ratio
indicator), because flux is measured consistently through the same aperture
across all bands, while the cModel magnitudes provide a more robust
estimate of the total flux based on independent model fits in each
bandpass~\citep[but see][]{Bernardi2017}. Therefore, we rescale the extinction-corrected {\it gri} model magnitudes to the $i$-band cModel magnitudes, and fit a Stellar Population
Synthesis~(SPS) model to the scaled {\it gri} magnitudes
to infer $\msbcg$.

\begin{figure*}
\centering\includegraphics[width=0.96\textwidth]{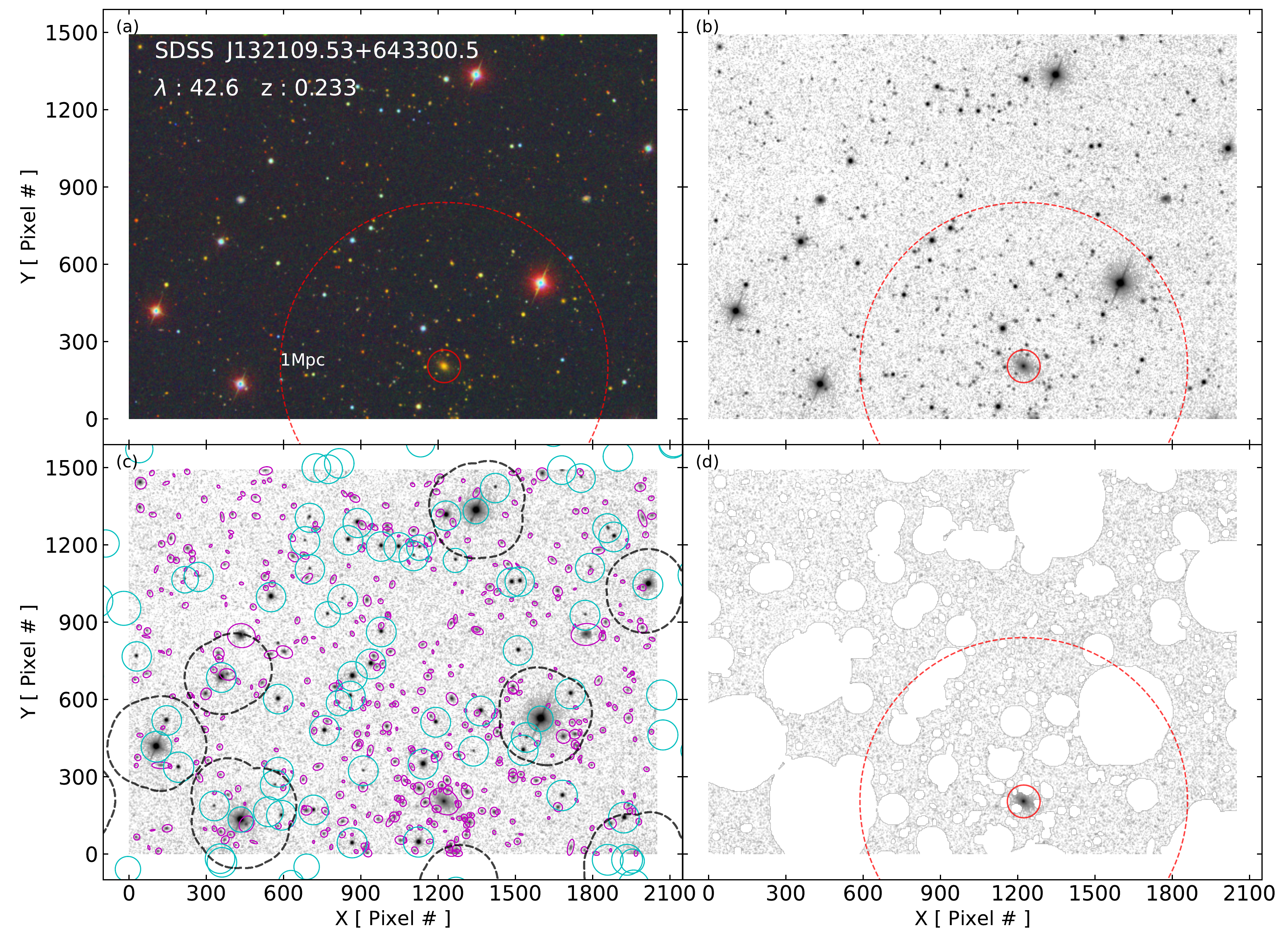}
\caption{Example of star and galaxy masks on the standard
    SDSS image frame ($589{\times} 811\,\sqarcsec$) of a cluster with
    $\lambda{=}42.6$ at $z=0.233$. The image dimension is
    $2048{\times}1489$ pixels, and the white margin area between the image
    frame and the panel edge has a width of $100$ pixels. Panel (a):
    SDSS $gri$-band composite image of the cluster, with its BCG indicated
    by the inner solid circle with a radius of $100\,\kpc$. The outer
    dashed circle indicates the cluster region of $1\,\mpc$ radius. Panel
    (b): The SDSS $r$-band image frame. Panel (c): Cyan circles and magenta
    ellipses are the star and galaxy masks, respectively, while the regions
    delineated by black dashed lines are the merged masks of saturated
    pixels surrounding bright stars. Note that we also include the masks of
    sources that are centred in the white margin area~(i.e., outside but
    close to the image frame), as they could still pollute the pixels
    inside the image frame. Panel (d): Final masked image that only
    includes the light from the BCG, ICL, and unmasked sources that are
    below the detection threshold. \label{fig:demo_mask} }
\end{figure*}

Following~\citet{Maraston2009}, we assume a simultaneous burst of two
Simple Stellar Populations~(SSPs)
at the same epoch, one dominant stellar population~(97 per cent) with solar
metallicity and the other a secondary~(3 per cent) metal-poor~($Z{=}0.008$)
population.  We utilize the \texttt{EzGal} software~\citep{Mancone2012} and
adopt the BC03 SSP model and the Chabrier IMF for the fits. For a detailed
comparison between our photometric stellar mass estimates and the
spectroscopic stellar masses from \citet{Chen2012}, we refer interested
readers to the Figure 1 in \citetalias{Zu2021}.

Due to the magnitude rescaling, our estimates of $\msbcg$ should inherit
the effective aperture of the $i$-band cModel magnitudes. By examining the
stacked surface stellar mass density profiles of clusters at fixed $\msbcg$
but different $\lambda$~(as will be demonstrated later in \S\ref{sec:soi}),
we find that the effective aperture of our $\msbcg$ estimates is
$R_*{\simeq}50\,\kpc$.  Therefore, our fiducial $\msbcg$ roughly
corresponds to $M_{*,50\kpc}$ in the language of~\citet{Huang2021}.  To
test the robustness of our $\rsoi$ detection, we also measure for each BCG
an aperture-based stellar mass $M_{*,20\kpc}$, i.e., stellar mass enclosed
within a fixed aperture of $20\,\kpc$.

\subsection{SDSS Images}
\label{subsec:img}

For any given set of clusters, we stack their SDSS images centred on the
BCGs and measure the average 1D SB profile from the stacked 2D image. In
this paper, we employ the observed images derived from the SDSS DR8, the
same imaging data from which the \redmapper~cluster catalogue was built.
The images were processed with the latest SDSS photometric pipeline
\texttt{photo v5\_6}, which implements an updated sky subtraction method
that significantly improves the flux estimates for bright objects,
detection of faint objects around bright objects, and extended light
measurement of large objects~\citep{Blanton2011}.  In particular, we make
use of the ``corrected frames'', i.e., the calibrated and sky-subtracted
images~(with bad columns and cosmic rays interpolated over) in the $g$,
$r$, and $i$ bands.  Each corrected frame has a dimension of 2048 pixels
${\times}$ 1489 pixels, which corresponds to a sky area of
$13.5{\times}9.8$ arcmin$^{2}$~(the pixel size is $0.396$ arcsec). The
improved photometric reduction of the SDSS images allows a robust
measurement of the large-scale, diffuse light distribution within massive
clusters, which was severely underestimated in the previous SDSS
photometric pipeline~\citep{Bernardi2013, Kravtsov2018}.

\begin{figure*}
    \centering\includegraphics[width=0.96\textwidth]{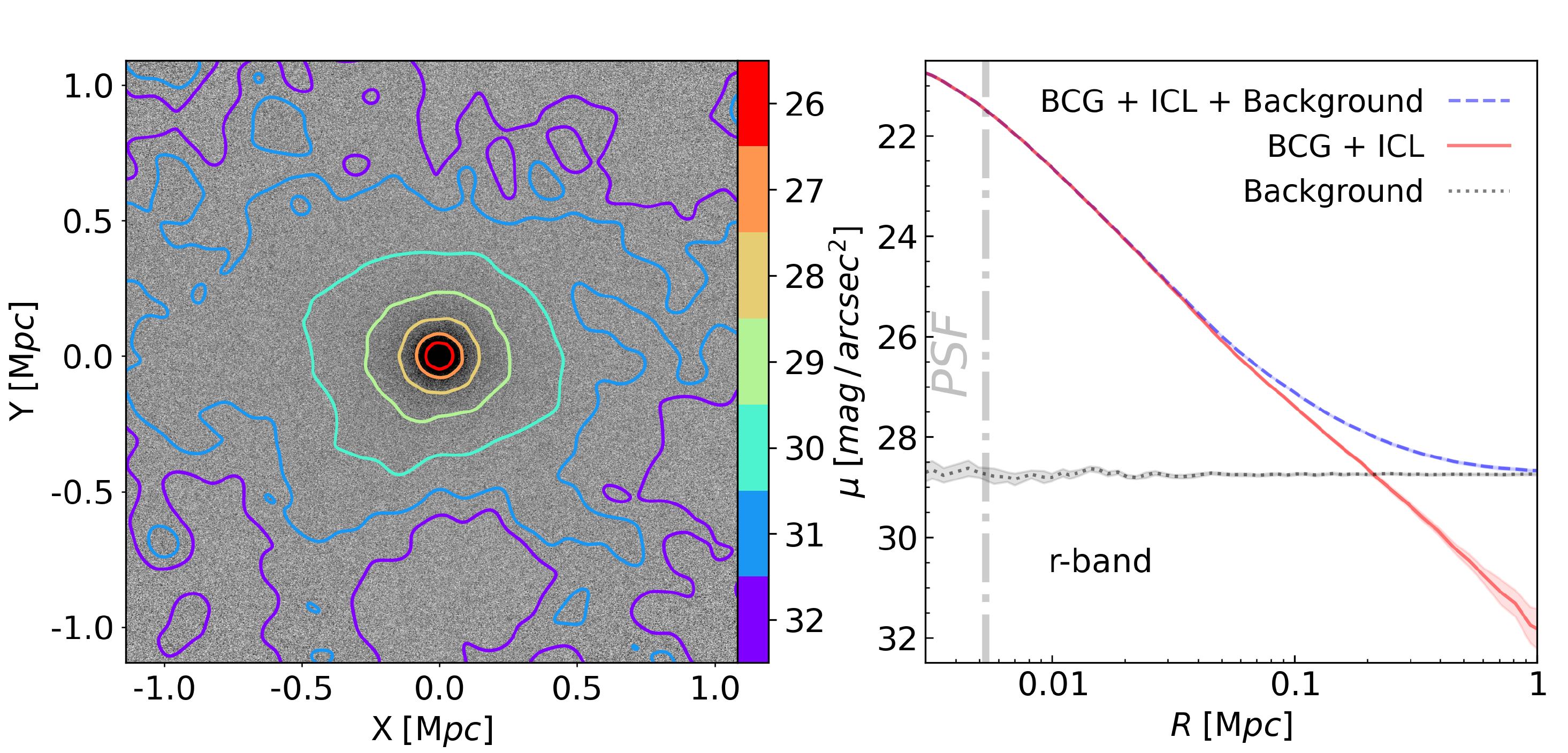}
    \caption{{\it Left}: $r$-band stacked image of the BCG+ICL light of our
    overall cluster sample~(grayscale). Contour lines indicate seven levels
    of SB ranging from $26$ to $32\,\sbmag$ with $\Delta\mu{=}1\,\sbmag$,
    colour-coded by the vertical colourbar on the right. Smoothing kernels
    of 3, 7, 11, 17, 29, 29, and 29 pixels are applied to the seven contour
    levels, respectively. {\it Right}: SB profiles of the BCG+ICL
    signal~(red solid), background~(gray dotted), and the sum of the signal
    and background~(blue dashed). Shaded band centred on each curve
    represents the $1\sigma$ uncertainties estimated from Jackknife
    resampling. Dot-dashed vertical line indicates the physical scale that
    the PSF size corresponds to at $z{=}0.25$.\label{fig:2D_all} }
\end{figure*}

We apply a two-step procedure to identify the defect images that
could severely corrupt the low-surface brightness signal. In the first
step, we perform visual inspection of each image to look for strong defects
that could severely undermine the stacked signal across all scales, e.g.,
extremely bright stars~(and their extended bright wings) and strong cosmic
rays across the frame. In the second step, we develop a simple
$\sigma$-clipping scheme to reject some mildly defective images that could
still impair our capability of detecting the faint ICL on scales larger
than a few hundred $\kpc$s. In particular, we divide each image
frame~(after masking out all the galaxies, bright stars, and saturated
pixels; described further below) into approximately 70
square patches with side length of 200 pixels, and measure the average pixel
flux ($\bar{f}_{\mathrm{pat}}$) within each patch. In addition, we measure
the mean~($\bar{f}_{\mathrm{img}}$) and scatter~($\sigma$) of the average
fluxes of the 70 patches, and compute the deviation of the average flux of
each patch from the mean flux of the entire image as $\Delta
f{\equiv}|\overline{f}_{\mathrm{pat}}{-}\overline{f}_{\mathrm{img}}|$. We
consider any image that includes a large block of $M$ connected patches
with $\Delta f{>}N \sigma$ to be a potential contaminant. After extensive
convergence tests, we find that the combination of $M{=}5$ and $N{=}6$
provides a robust selection criteria for culling out bright extended
features of non-cluster origin.  In total, we identify $1466$ strongly and
$1690$ mildly defective images of the $4593$ clusters from all three bands
and exclude them from further stacking analysis, leaving $2898$ clusters
within our final sample.

\section{Surface Brightness Profile}
\label{sec:sb}

To measure the SB profiles of the BCG+ICL $\mubi$, we mask out
the stars and non-BCG galaxies~(i.e., satellites and background
galaxies) from each image, and then transform all the images from the
cluster redshifts to the same reference redshift of
$z_{\mathrm{ref}}{=}0.25$~(with the same pixel size but different image
sizes).

Our method for image stacking is largely similar to that of
\citep[hereafter referred to as \citetalias{Zibetti2005}]{Zibetti2005}, but
with three important distinctions. Firstly, we directly utilize the
sky-subtracted corrected frames for our SB measurements, while
\citetalias{Zibetti2005} performed an independent sky subtraction on images
in the SDSS Early Data Release. We have tested the \citetalias{Zibetti2005}
sky-subtraction method on the raw DR8 SDSS images and find that the
\citetalias{Zibetti2005} method generally produces consistent results
compared to the SDSS pipeline.  Secondly, we estimate the background level
of the SB profiles, primarily due to undetected background galaxies and
foreground stars, by stacking the real SDSS images centred on the
coordinates of \redmapper~random clusters, while \citetalias{Zibetti2005}
inferred the background SB by extrapolating from the best-fitting projected
Navarro-Frenk-White~\citep[NFW;][]{NFW1997} profile of the SB
measurement below 900 $\kpc$. Lastly, we estimate the error matrix of our
SB profiles by resampling the cluster sample into 30 Jackknife
samples~(described further below), while \citetalias{Zibetti2005} divided
the annulus at each radius into $N$ angular sectors and calculated the
error on the mean SB as the rms of the $N$ SB values divided by
$\sqrt{N{-}1}$. We find that the diagonal errors from our Jackknife error
matrix is generally larger than the \citetalias{Zibetti2005} errors by a
factor of two on scales above $10\,\kpc$, likely because the SB of
neighboring sectors within the same annulus are highly correlated.  We
describe each step of our SB measurement in turn below.

\subsection{Image Masking}
\label{subsec:mask}

For any star brighter than $m_r{=}20$ and has the full width at half
maximum~(FWHM) measured in at least one of the other two bands, we adopt
the maximum FWHM of the three bands and multiply it by 30 as the diameter
of its mask. We further increase the diameter of the masks to
75$\times$FWHM for the saturated pixels. For stars detected only in $r$
band or fainter than $m_r{=}20$, we leave them unmasked in the image but will
remove their light statistically through background subtraction. We have
extensively tested the impact of different mask sizes on the measured
BCG+ICL SB profile, and verified that our choices yield the best
combination of signal-to-noise ratio~(S/N) and star light mitigation.

\begin{figure*}
    \centering\includegraphics[width=0.96\textwidth]{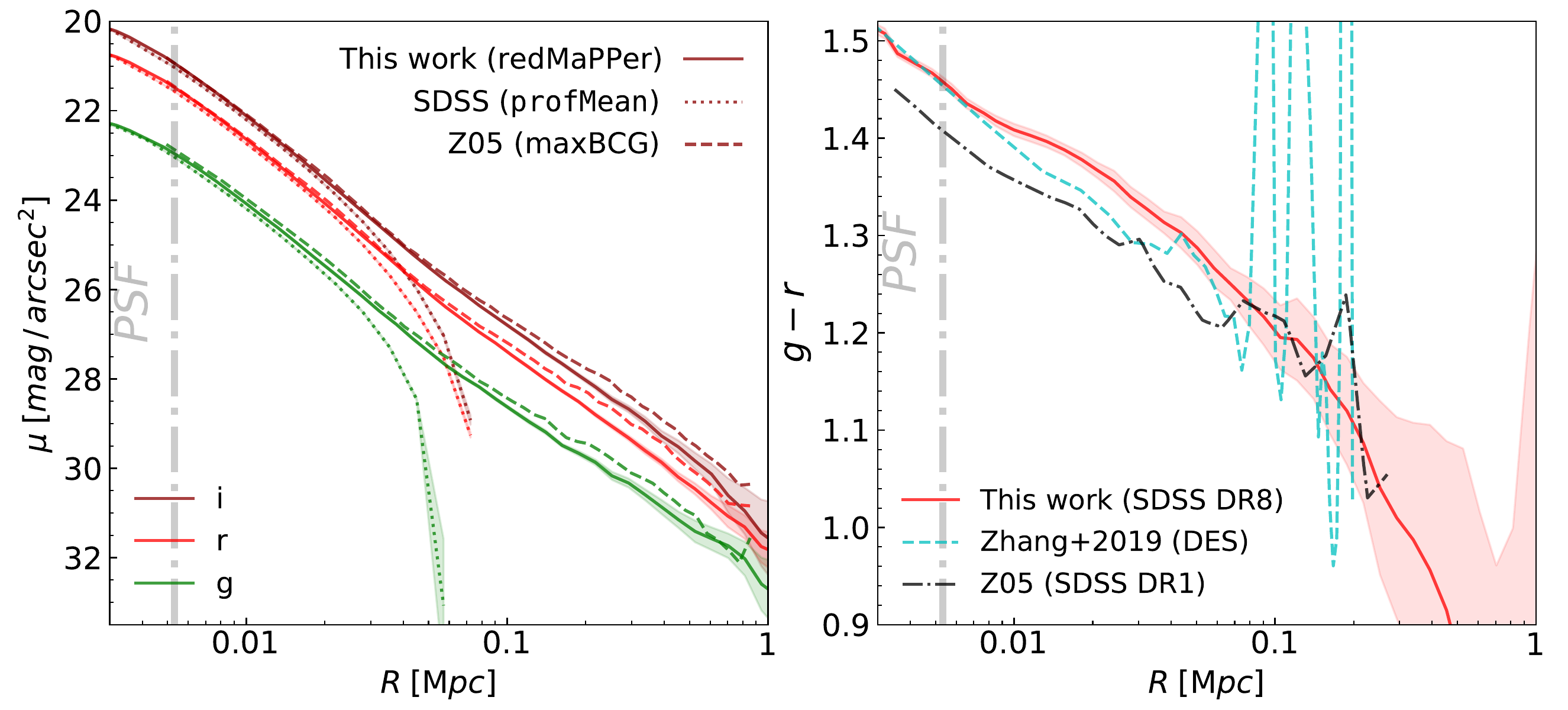}
    \caption{Comparison of our measurements of the BCG+ICL SB~(left) and
    colour~(right) profiles with previous studies. {\it Left}: Solid,
    dashed, and dotted curves indicate the BCG+ICL SB profiles from our
    measurements for \redmapper~clusters, \citetalias{Zibetti2005} for
    maxBCG clusters, and the SDSS photometric pipeline~(\texttt{profMean})
    for the \redmapper~BCGs, respectively. Green, red, and maroon curves
    are the measurements for the $g$, $r$, $i$ bands, respectively.  {\it
    Right}: Red solid, cyan dashed, and black dot-dashed curves indicate
    the BCG+ICL colour profiles measured by our work from SDSS DR8,
    \citetalias{Zibetti2005} from SDSS DR1, and \citet{Zhang2019} from
    DES, respectively.  In both panels, the dot-dashed vertical lines
    indicate the scale of the PSF, and the shaded bands around the solid
    curves represent the $1\sigma$ uncertainties estimated from Jackknife
    resampling.  \label{fig:Z05_comparison}}
\end{figure*}

For galaxy masks, we run \texttt{SExtractor}~\citep{Bertin1996} on the
cluster images in three bands separately, using a detection threshold of
$1.5\sigma$ and a minimum detection area of five pixels. Such detection
criteria result in a limiting magnitude of roughly $22.08$ magnitude in the
$i$ band~(i.e., an absolute magnitude of $-18.49$ for a galaxy at
$z{=}0.25$). We multiply the semi-major and semi-minor radii of the
ellipses detected by \texttt{SExtractor} both by a factor of eight, and
adopt the augmented ellipses as the masks for galaxies. We include the BCGs
in the masks for the image selection described in \S\ref{subsec:img}, but
leave them unmasked in the stacked profile measurements. The factor of
eight is a relatively conservative choice, which ensures that we remove
almost all the light from galaxy outskirts and the scattered
light from some of the very bright, nearby galaxies.

\begin{figure}
\begin{center}
\includegraphics[width=0.5\textwidth]{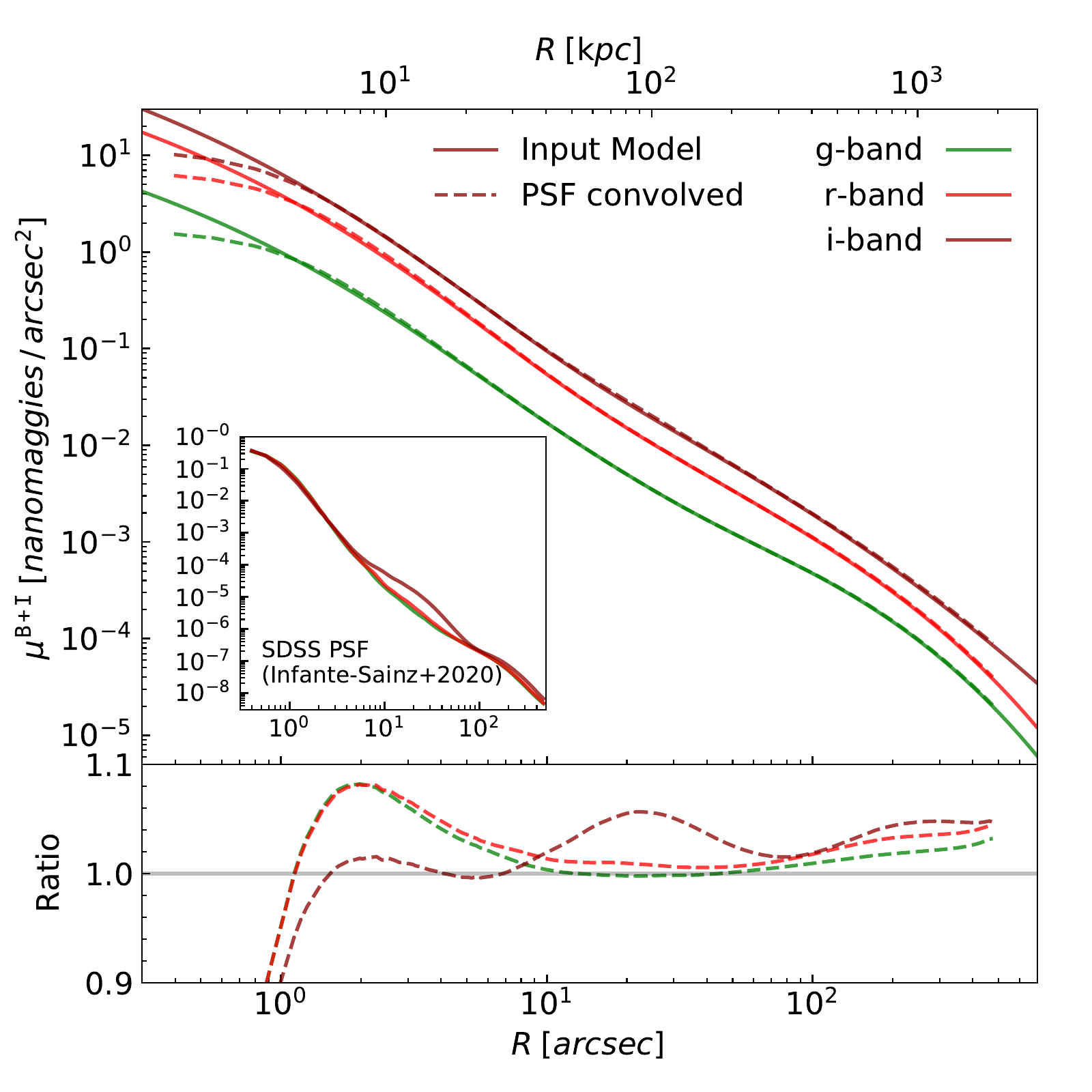}
\caption{Impact of the SDSS PSF on the BCG+ICL SB profiles.  {\it Top}:
Surface brightness profiles in the g~(green), r~(red), and i-band~(maroon).
Solid curves are the input three-S\'ersic models of the SB profiles before
convolving with the PSF, and the dashed curves are the results after
convolving them with the PSF derived by \citet{Infante-Sainz2020} in each
band~(illustrated in the inset panel, where the PSF in each band is
    normalized to have unit total flux). {\it Bottom}: The ratio between
    the SB profiles after and before PSF convolution. The bottom and top
    x-axes are in units of arc sec and $\kpc$~(physical), respectively.}
    \label{fig:SB_convo_psf}
\end{center}
\end{figure}

To further remove any contamination by bright stars and extended sources
from outside the image boundaries, we identify all the eight images
adjacent to each target image, and apply the same masking procedure to
those neighboring images.  We then merge the external star and galaxy masks
that overlap with the target image into the central mask. Finally, we
combine the three sets of masks from the {\it gri}-bands into a single
image mask, so that objects below the detection threshold in one particular
band but detected in another would still be masked out in that band. By
adopting a single uniform mask across three bands, we also ensure that the
measurement of colour profiles is robust against the discrepancy in the
masks of different bandpasses.

Figure~\ref{fig:demo_mask} demonstrates the efficacy of our masking
procedure, using the $r$-band corrected frame of a typical cluster in our
sample~(with $\lambda{=}42.6$ and $z{=}0.233$). Each panel has a dimension
of 2248 by 1689 pixels, larger than the original image frame by
200 pixels on each side~(i.e., the white strips surrounding each image
frame have a width of 100 pixels). Panel (a) shows the false-colour image
of the cluster, with the large dashed and small solid concentric circles
indicating 1 $\mpc$ and 100 $\kpc$-radius regions centred on the BCG,
respectively.  Panel (b) is similar to panel (a) but shows the $r$-band
image, with the grayscale indicating the individual pixel fluxes.  Panel
(c) shows the masks of the detected stars~(cyan solid circles), saturated
pixels~(black dashed circles), and galaxies~(magenta solid ellipses) within
the field. The circles within the white strips surrounding the original
image frame represent the sources from neighboring parts of the sky.
Clearly, some of the bright stars and saturated pixels in the white strips
could significantly pollute pixels of the cluster image. Panel (d)
shows the final $r$-band image after all the masks have been applied,
including those derived from the $g$ and $i$-band images. We expect the
fluxes within the cluster centre to be dominated by the BCG and ICL, but
there still exist some unmasked satellite galaxies, faint background
galaxies, and faint foreground stars in the final image of panel (d). We
will statistically remove the contamination in the BCG+ICL profiles by the
undetected~(hence unmasked) background sources in \S\ref{subsec:sbprofile}
and faint satellites in \S\ref{subsec:budget}.

\subsection{Image Rescaling to Reference Redshift}
\label{subsec:rescale}

To stack images at the same physical scale, we transform all the masked
images to the reference redshift $z_{\mathrm{ref}}{=}0.25$ with the same
pixel size. Firstly, we rescale the pixel size of each image by the square
of the angular diameter distance ratio between the observed and reference
redshifts, $(D_A^{\mathrm{obs}}/D_A^{\mathrm{ref}})^2$, and the flux within
the pixel by the square of the ratio between the two luminosity distances,
$(D_L^{\mathrm{obs}}/D_L^{\mathrm{ref}})^{2}$. Cosmic dimming is
automatically included as the pixel SB scales with
$[(1+z_{\mathrm{obs}})/(1+z_{\mathrm{ref}})]^4$. We do not apply any
$K$-correction to the cluster fluxes due to the lack of robust SED
templates for the ICL. However, since the reference redshift is close to
the median redshift of the sample, we anticipate that the amount of
$K$-correction in the stacked images should be largely cancelled out.
We explicitly test this using the colours of the BCGs to compute the
distribution of $K$-correction values in each band, and find that the mean
$K$-corrections are close to zero with a scatter of 0.025 magnitude in the
i-band, which contributes to a $1-2\%$ uncertainty in the stellar
surface density profiles. In addition, we correct for Galactic extinction
on the stacked images using the average $A_v$ of the BCG sightlines, rather
than the individual BCGs. Secondly, we resample all the rescaled images to
$0.396''$ per pixel, i.e., the original pixel size of the SDSS images.  We
then redistribute the image fluxes into the resampled pixels, so that the
flux in pixel $i$~(after resampling) is $f_i{=}\Sigma \mu_j A_{ij}$, where
$\mu_j$ is the SB of the $j$-th pixel in the rescaled image~(i.e., before
resampling) and $A_{ij}$ is the overlapped area between pixel $i$ and pixel
$j$.

\subsection{BCG+ICL SB Profile Measurement}
\label{subsec:sbprofile}

With all clusters shifted to the reference redshift, we now stack their
images centred on the BCGs in physical coordinates. We do not rotate the
cluster images, e.g., to align the BCGs by their major axes, as we are
mainly concerned with the 1D azimuthally-averaged profiles in this paper.
We carefully account for the masked regions and adopt the mean flux at each
pixel for the stacked image. A total SB profile $\mu^{\mathrm{tot}}(R)$ is
then computed as the azimuthally averaged SB within each annulus of radius
$R$. However, this total SB profile includes not only the light from the
BCG+ICL, but also contributions from the unmasked satellite galaxies,
background galaxies, foreground stars, and the Galactic cirrus~(which is
uncorrelated with the cluster positions on the sky), hence a
BCG+ICL+Background SB profile.
. We will estimate the unmasked satellite contribution by inspecting
the satellite stellar mass functions later in \S\ref{subsec:budget}, and
describe the removal of the background SB induced by the other three
components below.

\begin{figure}
    \includegraphics[width=0.48\textwidth]{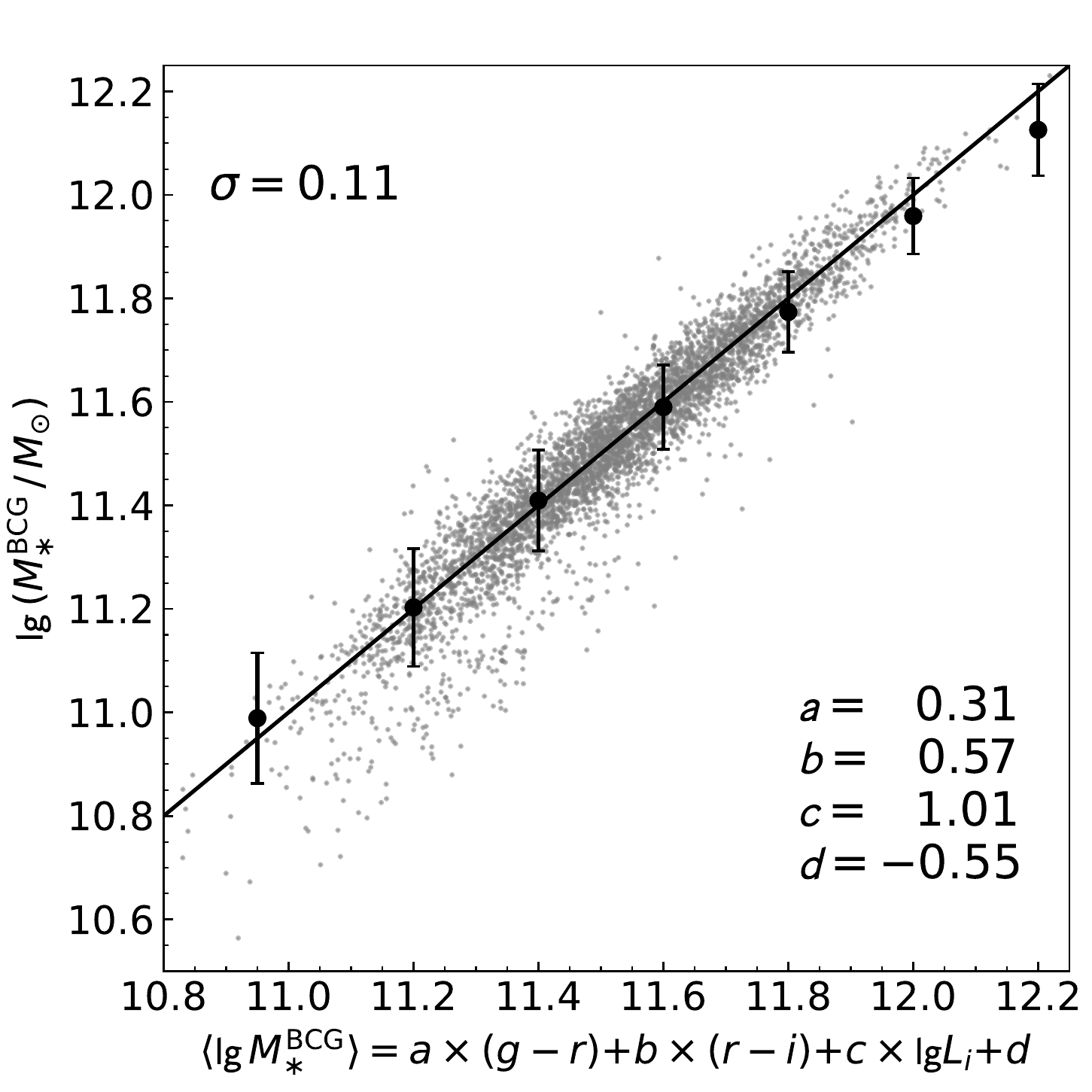}
    \caption{Comparison between the BCG stellar mass $\msbcg$ measured from
    template fitting~(y-axis) and that predicted by our linear $M_*/L_i$
    estimator~(x-axis). Black circles with errorbars indicate the median
    relation with its scatter~(${\sim}0.11$ dex), in good agreement with
    the one-to-one relation (black line). The parameters of the $M_*/L_i$
    estimator are listed in the bottom right corner.  \label{fig:func_M2L} }
\end{figure}

For any given set of clusters, we select a matching sample of random
clusters with the same joint distribution of redshift and richness from the
SDSS \redmapper~random catalogue v6.3.  Since the sensitivity of
\redmapper~algorithm to the background~(e.g., survey boundary, masks, and
variation of photometric depth) is incorporated in the generation of random
clusters~\citep{Rykoff2016}, we expect the background SB of the random
cluster images to be similar to that of the observed ones. To measure the
background SB profile, we download the SDSS corrected frames that host the
random clusters, and perform the same two-step quality inspection to remove
defect images from the random sample.  We then apply the same masking,
rescaling, and stacking procedures to the random cluster images that pass
the inspection, thereby producing a background SB profile for the clusters.
We repeat such procedure ten times by reshuffling the relative positions of
the random BCGs on the images after each measurement, and calculate the average of
the ten measurements as our the final background SB profile
$\mu^{\mathrm{bkg}}(R)$.

Finally, we derive the BCG+ICL SB profile $\mubi(R)$ by subtracting the
background SB profile from the total SB profile,
\begin{equation}
\mubi(R) = \mu^{\mathrm{tot}}(R) - \mu^{\mathrm{bkg}}(R).
\end{equation}
We further normalise the $\mubi(R)$ profile to be zero at projected
distance $R{=}2\,\mpc$, and focus on the SB signals at $R{<}1\,\mpc$ for
the rest of the paper. Note that the $\mubi(R)$ profile measured in this
way includes the contribution from the faint satellite galaxies that are
unmasked. We do not subtract this contribution from our measured $\mubi(R)$
and $\sigbi(R)$ profiles, but will nonetheless estimate the total stellar
mass of the unmasked satellites $\Sigma M_*^{\mathrm{unmasked}}$ in
\S\ref{subsec:budget}. In order to estimate the uncertainties of
$\mubi(R)$, we employ the standard Jackknife resampling technique by
dividing each cluster sample into 30 equal-size subsamples, and estimate
the error matrix from the 30 ``leave-one-out''~(i.e., using 29 subsamples
for each resampled measurement) experiments~\citep{Efron1981}.

\begin{figure*}
    \centering\includegraphics[width=0.96\textwidth]{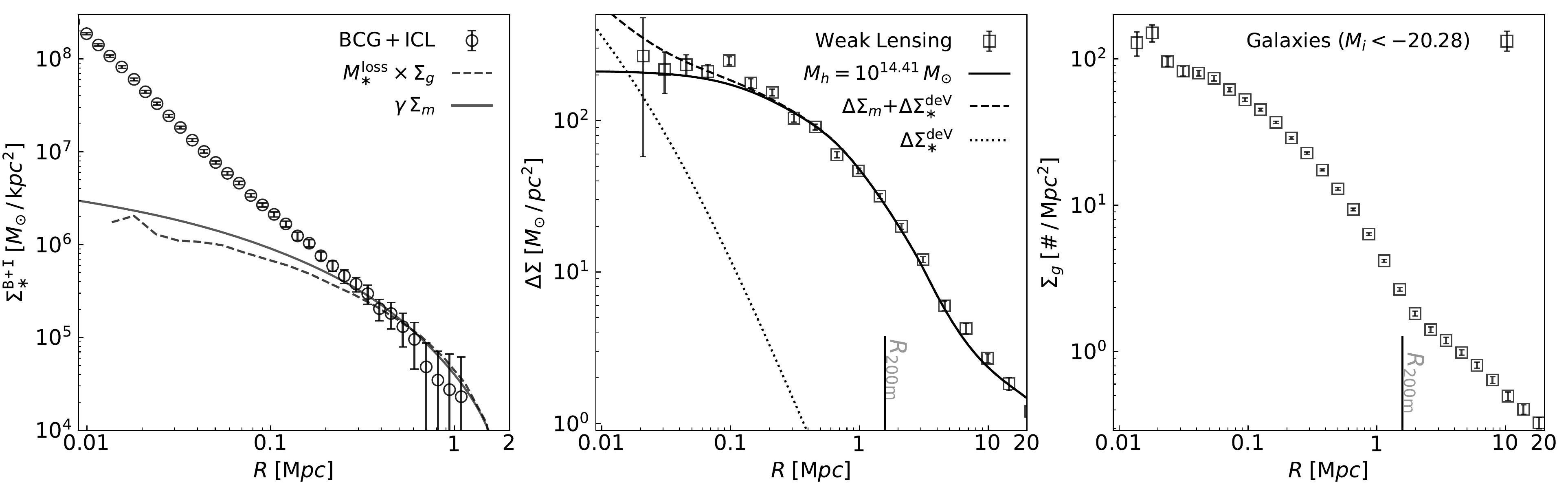}
    \caption{Distribution of stars, dark matter, and galaxies around the
    BCG.  {\it Left}: Stellar surface mass density profiles. Open circles
    with errorbars are the $\sigbi(R)$ profile measured by this work, while
    solid~($\gamma\sigm$) and dashed black~($M_*^{\mathrm{loss}}\sigg$)
    curves are the model predictions assuming that the outer ICL follows
    the distributions of dark matter~(middle panel) and satellite
    galaxies~(right panel), respectively. The best-fitting scaling factors
    are $\gamma{=}1/446$ and
    $M_*^{\mathrm{loss}}{=}1.4{\times}10^{10}\msol$.  {\it Middle}: Surface
    density contrast $\ds(R)$ measured by cluster weak lensing~(squares
    with errorbars) and predicted by the best--fitting model described in
    \citetalias{Zu2021}~(solid). Dotted curve indicates the weak lensing
    contribution from the stellar mass of the best-fitting de Vaucouleurs'
    profile, and dashed curve is the sum of the two contributions.  {\it
    Right}: Galaxy surface number density profile $\sigg(R)$ measured from
    the cluster-galaxy cross-correlation function. The vertical lines in
    the middle and right panels mark the halo radius $R_{200\mrm{m}}$.
    \label{fig:massprof3panel} }
\end{figure*}

Figure~\ref{fig:2D_all} shows the stacked 2D image of the BCG+ICL SB
distribution~(left) and the corresponding 1D SB profiles~(right) for our
overall cluster sample in the $r$ band. In the left panel, the grayscale
intensity represents the SB of each pixel, while the contour lines indicate
seven levels of SB ranging from $\mu{=}26\,\sbmag$ at $R{\simeq}50\,\kpc$
to $32\,\sbmag$ at $R{\simeq}1\,\mpc$~(with $\Delta \mu{=}1\,\sbmag$
increasing outwards),
obtained using smoothing kernels of 3, 7, 11, 17, 29, 29, and 29 pixels,
respectively and colour-coded by the vertical colourbar on the right.
The azimuthally averaged 1D SB profiles are shown in the
right panel, where the blue dashed, gray dotted, and red solid curves
indicate the total stacked SB $\mu^{\mathrm{tot}}(R)$~(BCG+ICL+background),
background SB $\mu^{\mathrm{bkg}}(R)$, and the BCG+ICL SB profiles
$\mubi(R)$, respectively. The dot-dashed vertical line indicates the
physical scale that the PSF~(point spread function) size corresponds to at
$z{=}0.25$ (the same as the dot-dashed vertical lines in both panels of
Figure \ref{fig:Z05_comparison}).  The shaded bands indicate the SB
uncertainties estimated from Jackknife resampling. Thanks to the large
sample size, we are able to robustly measure the diffuse cluster light down
to roughly $32\,\sbmag$ at $R{=}1\,\mpc$ despite the relatively shallow
depth of the SDSS imaging from a decade ago.

We compare our BCG+ICL SB~(left) and $g{-}r$ colour~(right) profiles with
the results from previous studies in Figure~\ref{fig:Z05_comparison}. In
the left panel, solid and dashed curves indicate the $\mubi(R)$ profiles
measured for the \redmapper~clusters in this work and the
maxBCG~\citep{Koester2007} clusters by \citetalias{Zibetti2005},
respectively, in the SDSS $g$~(green), $r$~(red), and $i$~(maroon) bands.
The amplitudes of the maxBCG profiles are slightly higher than the
\redmapper~ones on all scales, likely because the maxBCG clusters selected
by \citetalias{Zibetti2005} are on average higher mass systems than ours.
On small scales, our stacked profiles are consistent with the average BCG
light profiles~(dotted) derived by the SDSS photometric
pipeline~(\texttt{profMean}), but the dotted curves are rapidly cut off on
scales above $30\,\kpc$.  For the $g{-}r$ color profiles shown in the right
panel, our measurement~(red solid) is slightly redder than the that of
\citetalias{Zibetti2005}~(black dot-dashed), consistent with the
\citetalias{Zibetti2005} BCGs being more massive. We also show the DES
$g{-}r$ colour profile measured by \citet{Zhang2019}  for the DES
clusters~(cyan dashed), which exhibits a similar slope compared to
\citetalias{Zibetti2005} and our results. The \citeauthor{Zhang2019}
profile is measured from a much smaller cluster sample than ours with
${\sim}280$ DES \redmapper~clusters, so their measurements are cut off at
$200\,\kpc$ despite the DES photometry is roughly two magnitudes deeper
than SDSS.

Note that none of the BCG+ICL SB profile measurements shown in
Figure~\ref{fig:Z05_comparison} are corrected for the wide-angle effect of
the PSF~\citep{Tal2011, D'Souza2014, Wang2019}. To test the impact of PSF
on the diffuse light on large scales, \citet{Zhang2019} fit a model with
three S\'ersic components to the measured BCG+ICL SB profile, and convolve
the best--fiting model curve with the DES PSF in each band. They found that
the SB enhancement due to PSF on scales above $100\,\kpc$ is smaller than a
few percent and is thus negligible in the ICL analysis. Adopting the same
appproach, we fit a three S\'ersic model to the BCG+ICL SB profile in each
band, and convolve them with the SDSS 2D PSF model inferred by
\citet{Infante-Sainz2020}. Figure~\ref{fig:SB_convo_psf} compares the
difference between the SB profiles before~(solid) and after~(dashed) the
2D PSF convolution in the g~(green), r~(red), and i~(maroon) bands,
respectively. The inset panel illustrates the azimuthally-averaged SDSS 1D PSF inferred by
\citet{Infante-Sainz2020}, normalized so that total flux is unity in each
band, and the dashed curves in the bottom panel show the ratio profiles
between the SB profiles after and before 2D PSF convolution. In the g- and
r-bands, the
relative difference is well below $2\%$ and $5\%$ on scales between
$20-400\,\kpc$ and above $400\,\kpc$ at redshift 0.25, respectively. The
difference in the i-band is slight larger due to the bump feature at
$\sim$20 arcsec in the i-band PSF, but is still below $6\%$ on all scales
above $10\,\kpc$ and smaller than the statistical uncertainties on scales
above $100\,\kpc$.  We have also examined the impact of the 2D PSF on all the
other measurements in our paper, and find that our conclusions are
unaffected by the wide-angle effect of the SDSS PSF.

\section{Stellar Surface Density Profile}
\label{sec:sigma}

\subsection{Mass-to-Light Ratio}
\label{subsec:moverl}

For each cluster sample, we now convert the light profiles $\mubi$ measured
in three bands into a BCG+ICL stellar surface density profile $\sigbi$
using an empirically calibrated mass-to-light ratio against the $\msbcg$
measured in~\S\ref{subsec:cls}.  In particular, we assume the $i$-band
mass-to-light ratio~($M_*/L_i$) can be described by a simple linear
function of the $i$-band luminosity $L_i$, $g{-}r$, and $r{-}i$ colours,
\begin{equation}
    \lg(M_{\ast}/L_{i}) = a \cdot (g-r) + b \cdot (r-i) + (c - 1) \cdot \lg L_{i} + d.
    \label{eqn:moverl}
\end{equation}
We use $\msbcg$, i-band cModel magnitudes, and the model magnitude colours
of the BCGs to infer the values of $\{a, b, c, d\}$ via least-square
fitting. Figure~\ref{fig:func_M2L} demonstrates the efficacy of our
empirical calibration of $M_*/L_i$, where we show the distribution of BCGs
on the observed vs. predicted $\msbcg$ plane. Filled circles with errorbars
indicate the mean observed $\msbcg$ at fixed $\langle\msbcg\rangle$
predicted by Equation~\ref{eqn:moverl} with $\{a{=}0.31, b{=}0.57,
c{=}1.01, d{=}{-}0.55\}$, in good agreement with the one-to-one
relation~(solid line). The outliers in the bottom left corner mainly
consist of BCGs with relatively blue colours, which have minimal impact on
the least-square fit due to their small fraction.

However, since the outer ICL has a bluer colour than the inner BCG~(right
panel of Figure~\ref{fig:Z05_comparison}), extrapolating
Equation~\ref{eqn:moverl}, which is calibrated against the inner BCG, to
the outer ICL at $R{>}200\,\kpc$ may incur a systematic bias in the
$\sigbi$ measurement.  Fortunately, the bias should be sufficiently small
compared to our measurement uncertainties at those radii. For example, a
$10\%$ bias in the coefficient $a$ would result in a $0.03$ dex bias in
stellar mass for $g{-}r{\simeq}1$, much smaller than the $20\%$ statistical
uncertainty of $\sigbi$ at $R{\sim}200\,\kpc$~(as will be described further
below). Furthermore, our detection of $\rsoi$ relies on the difference
between the $\sigbi$ of two cluster subsamples, so the $M_*/L_i$-induced
bias in $\sigbi$ should be largely cancelled out.

\subsection{Surface Density Profiles of Stars, Dark Matter, and Galaxies}
\label{subsec:sigmathree}

In the left panel of Figure~\ref{fig:massprof3panel}, we apply our
best--fitting formula of $M_*/L_i$~(Equation~\ref{eqn:moverl}) to the $gri$
SB profiles in Figure~\ref{fig:Z05_comparison} and derive the BCG+ICL
stellar surface density profile $\sigbi(R)$ for the overall cluster sample,
as shown by the open circles with errorbars. Clearly, the $\sigbi$ profile
has a significant ICL component that extends to a few ${\times}10^4
\msol\kpc^{-2}$ at scales ${\sim}500\,\kpc{-}1\,\mpc$, where we expect that
the ICL largely follows the distributions of dark matter and satellite
galaxies. To compare the surface density profile of the diffuse component
with that of the dark matter and satellite galaxies, we show the
measurements of cluster weak lensing $\ds$ and galaxy number density
profile $\sigg$ for the overall sample in the middle and right panels of
Figure~\ref{fig:massprof3panel}, respectively~(squares with errorbars).

We obtain the $\ds$ and $\sigg$ measurements~(as well as the theoretical
model of $\ds$) by faithfully following the methods described in
\citetalias{Zu2021}.  Briefly, the surface density contrast profile
$\ds(R)$ is measured from weak lensing using the DECaLs DR8
imaging~\citep{Dey2019}, while the galaxy surface number density profile
$\sigg(R)$ is calculated by cross-correlating clusters with the photometric
galaxies within SDSS DR8. The photometric galaxies are selected with an
effective absolute magnitude limit of $M_i{=}-20.28$. We also show the
best-fitting $\ds_m(R)$ profile predicted by the theoretical model of
\citetalias{Zu2021} in the middle panel~(solid curve), which describes the
small-scale lensing using an NFW halo density profile~(with the cluster
miscentring effect constrained by X-ray observations), and the large-scale
lensing using a biased version of the matter clustering. From the $\ds$
modelling, we infer the average halo mass of our cluster sample to be
$\mh{=}2.57{\times}10^{14}\,\msol$, consistent with the results from
\citetalias{Zu2021}. The dotted curve indicates the $\ds_*^{\mathrm{deV}}$
signal from the stellar mass of the best-fitting de Vaucouleurs' profile at
$R{<}20\,\kpc$, and the dashed curve shows the sum of
$\ds_*^{\mathrm{deV}}$ and $\ds_m$. The vertical lines in the two panels
indicate the halo radius $R_{200m}$ inferred from the weak lensing mass.
The $\sigg(R)$ profile exhibits an inflection at $R_{200m}$, indicating the
familiar transition from the 1-halo term~(dominated by satellite galaxies)
to the 2-halo term in the halo occupation distribution
language~\citep[e.g.,][]{Zu2015}.  We refer interested readers to
\citetalias{Zu2021} for technical details that are beyond the scope of this
paper.

Returning to the left panel of Figure~\ref{fig:massprof3panel}.  Solid and
dashed curves show the matter surface density profile $\sigm$ and galaxy
surface number density profile $\sigg$, multiplied by a scale factor
$\gamma$ and the average mass loss per galaxy $M_*^{\mathrm{loss}}$,
respectively. In particular, $\sigm$ is related to $\ds$ via
\begin{equation}
    \ds(R) = \overline{\sigm}({<}R) - \sigm(R),
\end{equation}
where $\overline{\sigm}({<}R)$ is the average surface mass density within
$R$. The scaling factor is defined as
\begin{equation}
    \gamma(R) = \frac{\Sigma_{\mathrm{ICL}}(R) +
    \Sigma^{\mathrm{unmasked}}_{\mathrm{sat}}(R)}{\sigm(R)}
    \label{eqn:gamma}
\end{equation}
where $\Sigma_{\mathrm{ICL}}$ is the ICL stellar mass profile and
$\Sigma^{\mathrm{unmasked}}_{\mathrm{sat}}$ is the stellar mass profile of
the unmasked satellite galaxies. By assuming that the ICL and unmasked
satellites both follow the distribution of dark matter, the above equation can be simplified as
\begin{equation}
    \gamma(R)\equiv \gamma =f_{\mathrm{ICL}} + f^{\mathrm{unmasked}}_{\mathrm{sat}},
    \label{eqn:gamma2}
\end{equation}
where $f_{\mathrm{ICL}}$ and $f^{\mathrm{unmasked}}_{\mathrm{sat}}$ are the
ICL and unmasked satellite stellar-to-halo mass fractions, respectively.  We
infer the best-fitting values of $\gamma{=}1/446$ and
$M_*^{\mathrm{loss}}{=}1.4{\times}10^{10}\,\msol$ by matching the scaled
profiles to the $\sigbi(R)$ measurements above $R{=}400\,\kpc$. We will
infer the values of $f_{\mathrm{ICL}}$ and
$f_{\mathrm{unmasked}}^{\mathrm{sat}}$ separately in \S\ref{subsec:budget}.
Note that we predict the $\sigm$ profile from the best-fitting $\ds$ model
curve in the middle panel, and directly adopt the observed $\sigg$ profile
in the right panel.  Additionally, both curves are normalized to have zero
amplitudes at $R{=}2\,\mpc$, following the same practice when measuring
$\sigbi$.

Overall, the two scaled profiles $\gamma\sigm$ and
$M_*^{\mathrm{loss}}\sigg$ in the left panel of
Figure~\ref{fig:massprof3panel} provide good descriptions to the BCG+ICL
stellar surface density profiles at $R{>}400\,\kpc$, indicating that the
ICL in the outer region of clusters indeed follows the distribution of the
dark matter and satellite galaxies. This is consistent with the findings
from \citet{Montes2019}, which suggested that the ICL is an excellent
tracer of dark matter, as well as satellite stripping being the dominant
channel of ICL production in the outer region. In particular, we find that
our observation of the diffuse light on scales above $400\,\kpc$ can be
explained if $\sim$0.1{-}0.2\% of the total mass is in the form of ICL, and
if ${\sim}10^{10}\msol$ of stars were stripped from each satellite galaxy
into the ICL.

\subsection{Decomposition of the BCG+ICL Surface Stellar Mass Profile}
\label{subsec:decomposition}

\begin{figure}
    \includegraphics[width=0.48\textwidth]{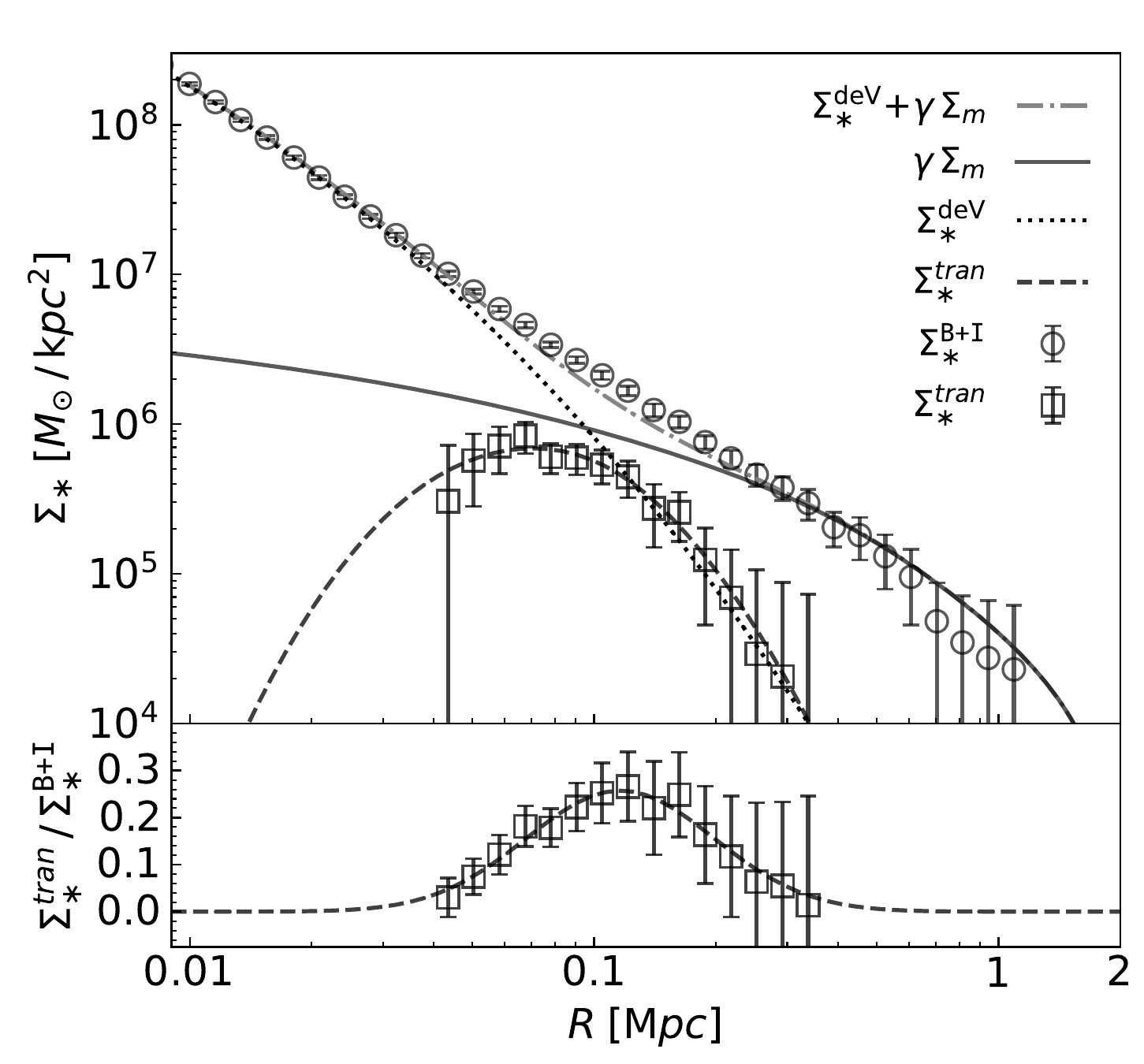}
    \caption{Decomposition of the BCG+ICL stellar surface mass density
    profile $\sigbi(R)$~(open circles with errorbars). Solid curve is the
    total surface mass density profile $\sigm$, scaled by
    $\gamma{\equiv}1/446$~(Equation~\ref{eqn:gamma}), while the
    best-fitting de Vaucouleurs profile for the $\sigbi$ profile at
    $R{<}20\,\kpc$ is shown as the dotted curve.  The sum of $\gamma\sigm$
    and $\sigdev$ is indicated by the dot-dashed curve, revealing an excess
    mass in the observed $\sigbi$ profile on transitional scales of
    $R{\sim}70{-}200\,\kpc$. Open squares with errorbars indicate this
    transitional component $\sigtr$. The bottom subpanel shows the ratio
    between $\sigtr(R)$ and $\sigbi(R)$, which can be described by a
    simple Gaussian~(dashed; Equation~\ref{eqn:sigtr}).\label{fig:decomposition} }
\end{figure}

Given that the outer region of the ICL roughly follows the distribution of
dark matter, we can separate the observed BCG+ICL stellar surface mass
profile $\sigbi$ into at least two physically distinct components,
including one that follows the dark matter on large
scales~($R{=}400\,\kpc{-}1\mpc$) and the other BCG-dominated portion on small
scales~($R{<}50\,\kpc$). By further assuming that the {\it intrinsic} BCG
can be described by a de Vaucouleurs' profile, the BCG-dominated portion
may include a third component on transitional scales where the extended BCG
envelope unfolds into the ICL. Following this philosophy, we can decompose
the $\sigbi$ profile via the following three steps. Firstly, we adopt the
total surface mass profile $\sigm$ inferred from weak lensing, and
multiply it by $\gamma{=}1/446$ to
describe $\sigbi$ at $R{>}400\,\kpc$, as was done in the left panel of
Figure~\ref{fig:massprof3panel}.
\begin{equation}
    \gamma \sigm(R) = \Sigma_*^{\mathrm{ICL}}(R)  +
    \Sigma_{\mathrm{sat}}^{\mathrm{unmasked}}(R)
    \label{eqn:sigmaicl}
\end{equation}
Secondly, we fit a de Vaucouleurs' profile to the $\sigbi$ measurement on
scales below $R{=}20\,\kpc$,
\begin{equation}
    \sigdev(R) = \Sigma_{e} \exp \left( -7.676 \left[\left(R/R_{e}\right)^{1/4} -
    1\right]\right),
\label{eqn:sigmabcg}
\end{equation}
yielding $\Sigma_{e}{=}10^{7.93}\msol/\kpc^2$ and $R_{e}{=}15.06\,\kpc$. Finally, we subtract
$\gamma \sigm$ and $\sigdev$ from the measured $\sigbi$ profile,
leaving us a ``transitional component''
\begin{equation}
    \sigtr(R) = \Sigma_*^{\mathrm{BCG+ICL}}(R) -
    \left[\Sigma_*^{\mathrm{deV}}(R) + \gamma \sigm(R)\right].
    \label{eqn:sigmatran}
\end{equation}
Therefore, $\sigtr(R)$ represents the excess component in the diffuse light
that cannot be described by the sum of a de Vaucouleurs' profile and an ICL
mass profile that follows the dark matter.

Figure~\ref{fig:decomposition} illustrates the physical decomposition of
our observed $\sigbi$ profile~(open circles with errorbars) into three
distinct components: a de Vaucouleurs' profile $\sigdev$~(dotted curve), a
scaled dark matter profile~$\gamma\sigm$~(solid curve), and a transitional
component~$\sigtr$~(open squares with errorbars).  The errorbars of
$\sigtr$ are inherited from that of the $\sigbi$ measurement, assuming zero
uncertainties from the subtraction. Additionally, dot-dashed curve indicates
the sum of the de Vaucouleurs' profile and the scaled dark matter profile,
which clearly under-predicts the signal on scales between $50\,\kpc$ and
$300\,\kpc$ --- a third component $\sigtr$ is required to fully describe
the diffuse light.  The ratio profile between $\sigtr(R)$ and $\sigbi(R)$
is shown in the bottom panel of Figure~\ref{fig:decomposition}. The
transitional component accounts for about 17\% of the total diffuse
mass on scales between $50\,\kpc$ and $300\,\kpc$, and the ratio peaks
at 25\% around $100\,\kpc$.

To test the robustness of $\sigtr(R)$, we fit the de Vaucouleurs' profile
to a larger maximum radius of $30\,\kpc$, finding that the centroid and
amplitude of $\sigtr(R)$ stay almost unchanged.  We also perform the
decomposition using a generic S\'ersic profile instead of a de Vaucouleurs'
profile, which has a fixed S\'ersic index of four.
Although the peak amplitude reduces from $25\%$ to $15\%$ once we allow the
S\'ersic index to deviate from four, the centroid stays unchanged at
${\sim}100\,\kpc$ and the $\sigtr(R)$ component remains significant
compared to the $\sigbi$ uncertainties~(${\sim}5\%$ at $R{=}100\,\kpc$).
Therefore, we conclude that the transitional component is required to
provide a complete description of the diffuse stellar mass profile on
scales between $50\,\kpc$ and $300\,\kpc$, regardless of assumptions for
the stellar mass profile of the inner BCG. For the rest of the paper, we
will adopt the $\sigtr(R)$ profile calculated from assuming a de
Vaucouleurs' profile at $R{<}20\,\kpc$ as our fiducial model of the
transitional component.

Finally, the ratio profile between $\sigtr$ and $\sigbi$ (shown in the
bottom panel of Figure \ref{fig:decomposition}) can be described by a
Gaussian~(in log-space):
\begin{equation}
\frac{\sigtr(R)}{\sigbi(R)} =  f_{t} \exp \left[ - \frac{(\lg R-\lg
	R_t)^2}{2\sigma_t^2}\right] ,
\label{eqn:sigtr}
\end{equation}
where $f_t{=}0.26$, $\sigma_t{=}0.23$ dex, and $R_t{=}116.0\,\kpc$ are the
amplitude, characteristic log-width, and centroid of the transitional
component, respectively.  The best-fitting function is shown as the
dashed curves in both panels of Figure \ref{fig:decomposition}.

\begin{figure}
    \includegraphics[width=0.48\textwidth]{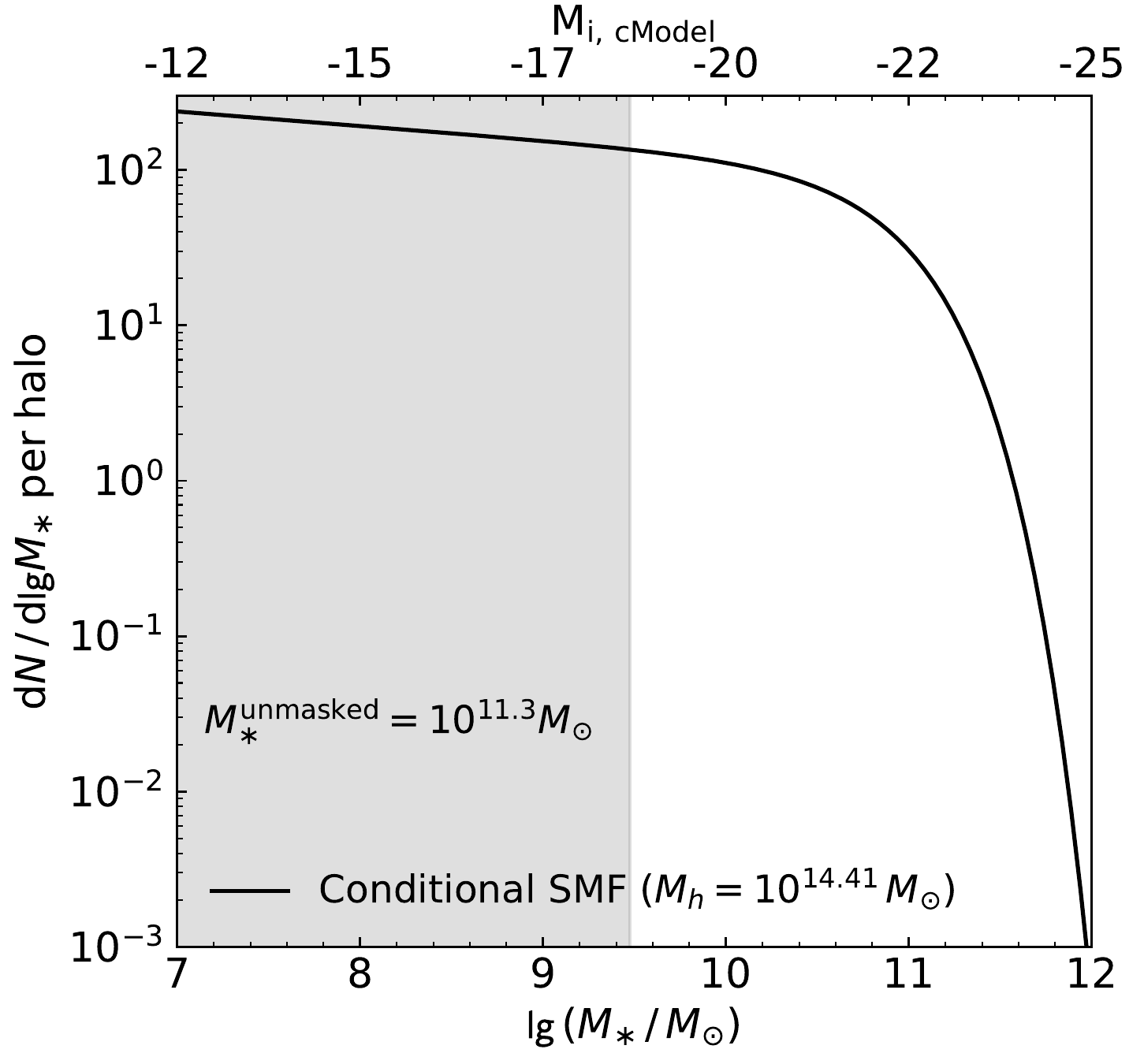}
    \caption{Conditional satellite stellar mass function of our cluster
    sample~(solid curve). Gray shaded region indicates the stellar mass
    range of satellites that are undetected by our source finding
    algorithm~(hence unmasked). The top x-axis shows the corresponding
    $i$-band cModel absolute magnitudes~(assuming at the reference redshift
    $0.25$). The total stellar mass of the unmasked satellite galaxies is
    $M_*^{\mathrm{unmasked}}=2{\times} 10^{11}\msol$. \label{fig:csmf} }
\end{figure}

\subsection{Stellar Mass Budget of Clusters}
\label{subsec:budget}

\begin{figure}
    \includegraphics[width=0.48\textwidth]{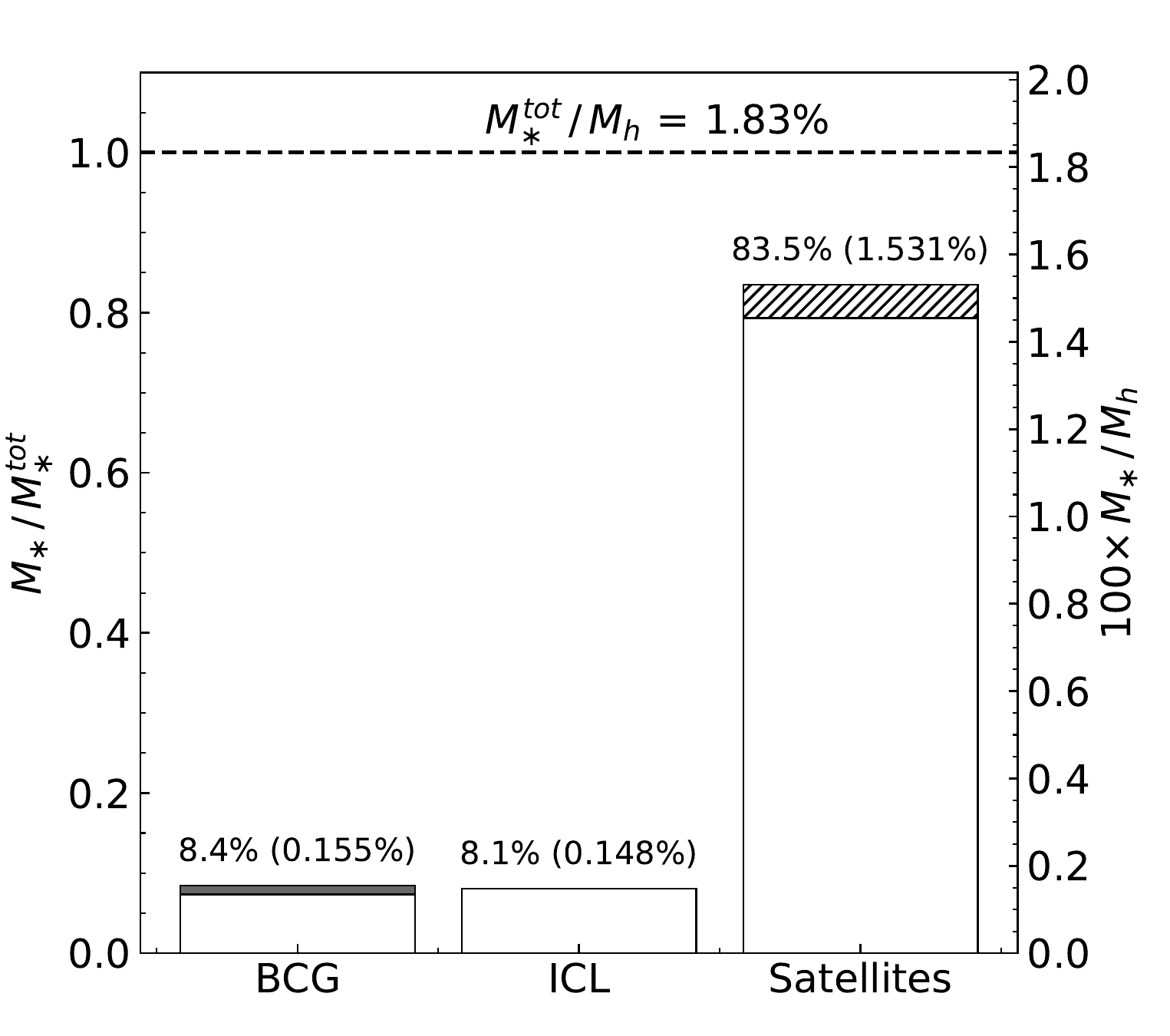}
    \caption{Stellar mass budget of our cluster sample in BCG, ICL, and
    satellites. Gray filled and hatched portions indicate the contributions
    from the transitional component and the unmasked satellite galaxies,
    respectively. The stellar mass fraction~(left y-axis) and
    stellar-to-halo mass fraction~(right y-axis) of each component are
    marked by the percentage values out and inside the parentheses,
    respectively. The horizontal line indicates the total stellar-to-halo
    mass fraction of the clusters $M_*^{\mathrm{tot}}/M_h$.
    \label{fig:budget} }
\end{figure}

With the physical decomposition in Figure~\ref{fig:decomposition}, we are
now ready to derive the stellar mass budget of clusters in three states:
BCG~(i.e., de Vaucouleurs + transitional), ICL, and satellite galaxies. In
order to estimate the total amount of stellar mass inside the satellite
galaxies, we make use of the conditional stellar mass function~(CSMF) of
clusters inferred by \citet{Yang2012}.  In particular, we adopt the shape
of the measured CSMF for their halo mass bin of
$10^{14.2}{-}10^{14.5}\,\hhmsol$~\citep[c.f., column 9, table 7
of][]{Yang2012}, which can be described by a Schechter function
\begin{equation}
    \frac{\dd N(M_* | M_h)}{\dd\lg M_*} = \Phi^{\star} \left( \frac{
    M_*}{M^{\star}} \right)^{\alpha+1} \exp \left( -
  \frac{M_*}{M^{\star} } \right)
    \label{eqn:csmf}
\end{equation}
with $M^{\star}{=}8.32{\times}10^{10}\msol$ and $\alpha{=}{-}1.093$.
However, the \citeauthor{Yang2012} CSMF was measured for a sample of galaxy
groups at $z{<}0.1$, which has a different value of $\Phi^{\star}$ than
that of our cluster sample at $z{\sim}0.25$. To determine $\Phi^{\star}$,
we obtain the number of satellite galaxies above our $i$-band absolute
magnitude limit of $-20.28$ to be $60$ per halo, by integrating the galaxy
surface number density profile $\sigg$ to
$r_{200\mrm{m}}{=}1.58\,\mpc$~(right panel of
Figure~\ref{fig:massprof3panel}). Given that the $i$-band magnitude of
$-20.28$ roughly corresponds to $\lg\,M_*{=}10.2$~(assuming a $M_*/L_i$ of
$1.88$), we normalise Equation~\ref{eqn:csmf} for our cluster sample by
enforcing the total number of satellites above $\lg\,M_*{=}10.2$ to be $60$
per halo, yielding a $\phi^{\star}{=}79.8$ galaxies per dex in $M_*$ per
halo.  Figure~\ref{fig:csmf} shows the correctly normalized CSMF of our
cluster sample~(solid curve), which we integrate across the entire stellar
mass range to obtain the average stellar mass of the satellites
$\langle{M_*^{\mathrm{sat}}}\rangle{=}1.34{\times} 10^9\,\msol$ and the
total amount of satellite stellar mass $\Sigma
M_*^{\mathrm{sat}}{=}3.94{\times}10^{12}\,\msol$.

\begin{figure*}
    \centering\includegraphics[width=0.98\textwidth]{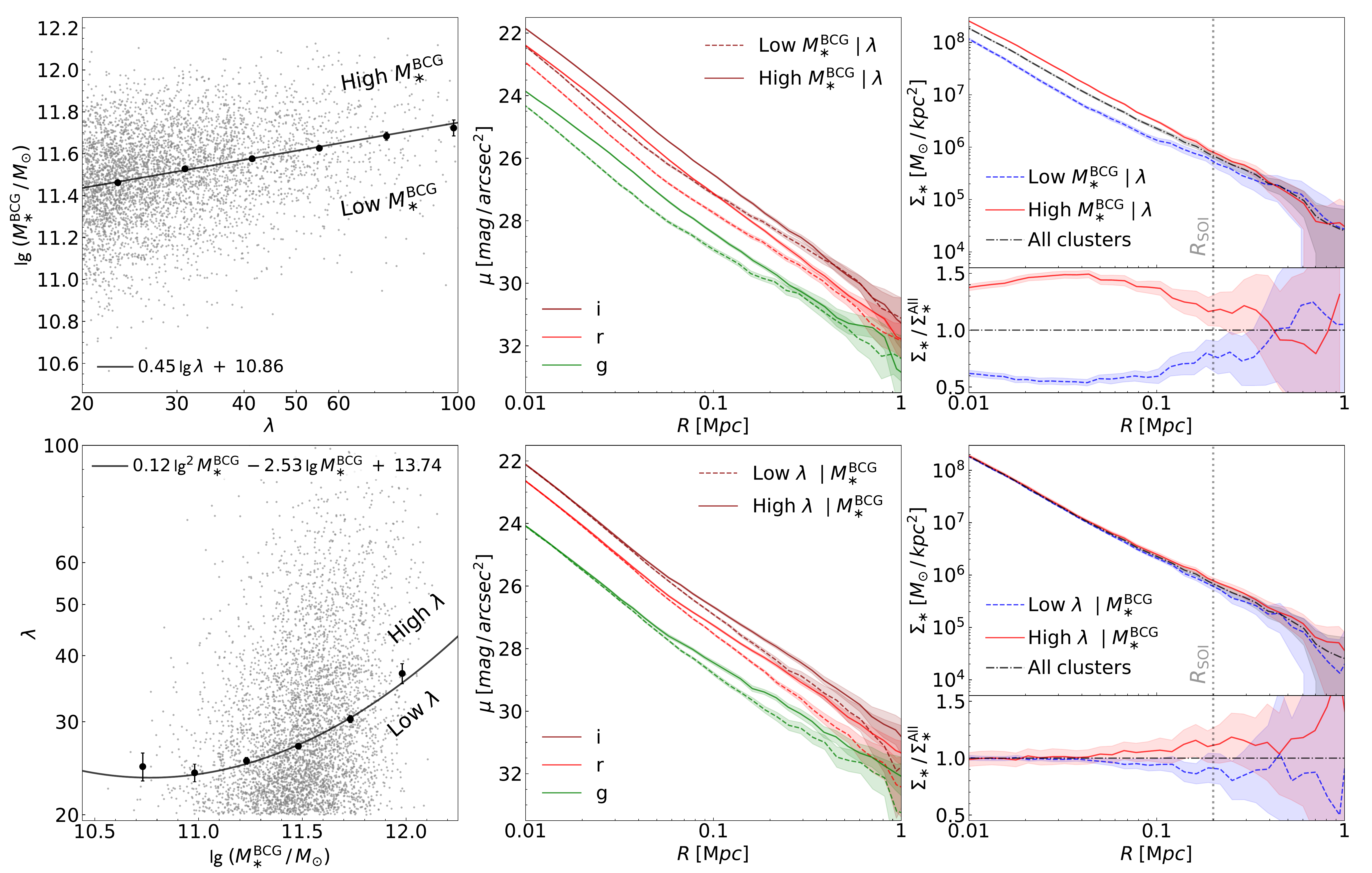}
    \caption{Distributions of clusters on the $\lambda$ vs. $\msbcg$
    plane~(left), surface brightness profiles~(middle), and stellar surface
    density profiles~(right) of cluster subsamples split by $\msbcg$~(top
    row) and $\lambda$~(bottom row).
	{\it Top left}: Circles show the median
    of $\lg\msbcg$ at fixed $\lg\lambda$, and black solid line is a linear
    fit to median relation, which divides the clusters into low and
    high-$\msbcg$ subsamples. {\it Top middle}: Surface brightness profiles
    $\mubi$ of the low~(dashed curves) and high~(solid curves) $\msbcg$
    subsamples. Green, red, and maroon curves indicate the measurements
    from $g$, $r$, and $i$-band images, respectively. {\it Top right}:
    Stellar surface density profiles $\sigbi$ of the low-$\msbcg$~(blue
    dashed), high-$\msbcg$~(red solid), and all~(gray dot-dashed) clusters.
    The bottom subpanel shows the ratios of the $\sigbi$ profiles of the
    low~(blue dashed) and high~(red solid) $\msbcg$ subsamples over that of
    the overall sample. Dotted vertical line indicates $\rsoi{=}200\,\kpc$.
    The panels in the bottom row are similar to those in the top, but for
    subsamples split by $\lambda$ at fixed $\msbcg$.\label{fig:split}}
\end{figure*}

Furthermore, the gray shaded region~(below $\lg\,M_*{=}9.48$) in
Figure~\ref{fig:csmf} indicates the stellar mass range that is below the
detection threshold of our source finding algorithm, hence unmasked during
the SB measurement.  This detection threshold roughly corresponds to the
i-band limiting cModel magnitude of $22.08$, i.e., an absolute magnitude of
$-18.49$ for a galaxy at $z{\sim}0.25$, yielding a total unmasked stellar
mass of $M_*^{\mathrm{unmasked}}{=}2{\times}10^{11}\msol$. We remove
this unmasked stellar mass contribution from the ICL stellar mass budget as
follows.  The unmasked satellite mass fraction is
$f^{\mathrm{unmasked}}_{\mathrm{sat}}{=}\Sigma
M_*^{\mathrm{unmasked}}/M_h{=}0.0762\%$. Since the scale factor defined in
Equation~\ref{eqn:gamma2} is $\gamma{=}1/446$, the ICL mass
fraction is $f_{\mathrm{ICL}}{=}\gamma{-}f^{\mathrm{unmasked}}_{\mathrm{sat}}
{\simeq}1/675$.

Finally, Figure~\ref{fig:budget} shows the stellar mass budget of our
cluster sample. We show the stellar mass fractions of the BCG, ICL, and
satellites in the left y-axis, and the stellar-to-halo mass fractions of the
three components in the right y-axis.  The gray filled portion on top of
the ``BCG'' histogram indicates
the integrated mass within the $\sigtr$ profile~($5.3{\times}10^{10}\,\msol$)
shown in Figure~\ref{fig:decomposition}, while the hatched region
on top of the ``Satellites'' histogram
corresponds to the total stellar mass of the unmasked satellites in Figure~\ref{fig:csmf}. The dashed
horizontal line in the top indicates the total stellar mass-to-halo mass
fraction~($1.83\%$) assuming a halo mass of $2.57{\times}10^{14}\,\msol$
measured from weak lensing in \citetalias{Zu2021}. Assuming further that
the total baryon fraction of the clusters is the cosmic value
$f_{b}{=}\Omega_{b}/\Omega_{m}{\simeq}16\%$, we can infer that
${\sim}88\%$ of the baryons are in the form of hot gas within clusters.
For the stellar mass budget, $83.5\%$ of the stellar mass is inside the
satellite galaxies and $8.4\%$ is in the BCG, leaving $8.1\%$ of the
stellar mass in the diffuse form of free-floating ICL stars --- hence
stellar-to-halo mass ratios of $1.531\%$~(satellites), $0.155\%$~(BCG), and
$0.148\%$~(ICL). Using the DES imaging, \citet{Zhang2019} estimated that
the luminosity fraction
of BCG+ICL is $44\pm17\%$, but they did not subtract the contribution from
undetected satellites. In our stellar mass budget, the sum of the stellar fraction
of the BCG+ICL and undetected satellites is about 21\%, comparable to the
lower end of the \citet{Zhang2019} estimation.

\section{BCG Sphere of Influence: Detecting \texorpdfstring{$\rsoi$}{Rsoi}}
\label{sec:soi}

Although we tentatively assign the transitional component to the BCG in
Figure~\ref{fig:budget}, it is unclear whether this stellar mass excess is
primarily tied to the BCG or the richness. Despite accounting
only for ${\sim}1\%$ of the total stellar mass, this transitional component is
key to solving the sphere of influence of the BCGs $\rsoi$. In particular,
if the excess mass is primarily richness-induced, we expect $\rsoi$ to stop
at $R_t{\times}10^{-\sigma_t}{\simeq}70\,\kpc$; but a BCG-induced origin
would extend $\rsoi$ beyond the transitional component at
$R_t{\times}10^{\sigma_t}{\simeq}200\,\kpc$. In this section, we divide our
overall cluster sample into two subsamples of different average BCG stellar
mass $\msbcg$ at fixed satellite richness $\lambda$, and aim to distinguish
the two physical scenarios by comparing the two sets of diffuse light and
mass profiles.

\subsection{Cluster Subsamples Split by \texorpdfstring{$\msbcg$}{Mstar}}
\label{subsec:split}

\begin{figure*}
    \centering\includegraphics[width=0.96\textwidth]{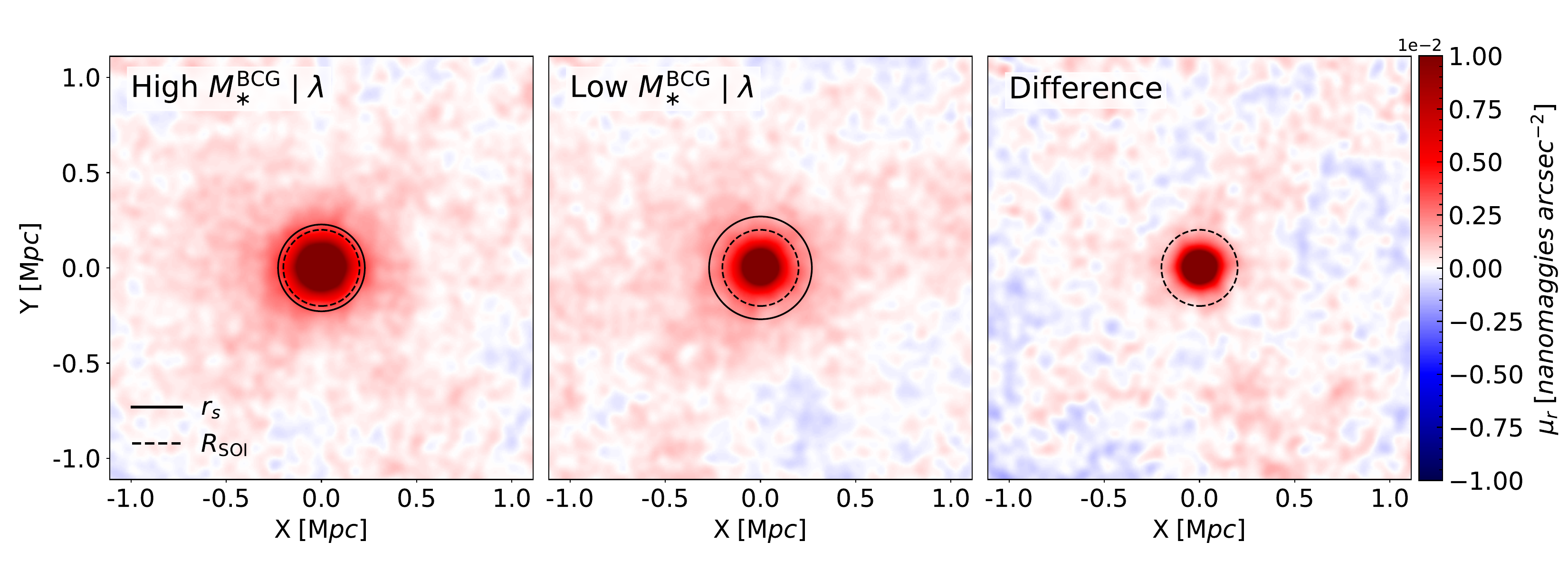}
    \caption{{\it Left}: Stacked 2D image of the high-$\msbcg$ clusters,
    with the pixel SB colour-coded by the vertical colourbar on the right.
    Solid and dashed circles indicate the cluster characteristic
    radius~($r_{s}{=}227.8\,\kpc$) and BCG sphere of
    influence~($\rsoi{=}200\,\kpc$), respectively.  {\it Middle}: Similar
    to the left panel, but for the low-$\msbcg$ clusters with
    $r_{s}{=}269.7\,\kpc$. {\it Right}: The difference image between the
    high and low-$\msbcg$ clusters. Clearly, the region outside
    $\rsoi$~(dashed circle) is largely consistent with zero. A smoothing
    kernel of 15 pixels is applied to the three images.  \label{fig:image2Dsub}}
\end{figure*}

Following \citetalias{Zu2021}, we split the clusters into two subsamples
using the median $\msbcg$-$\lambda$ relation, illustrated by the top left
panel in Figure~\ref{fig:split}. In particular, the median
$\msbcg$-$\lambda$ relation~(solid line) can be described by
\begin{equation}
\langle \lg\msbcg \rangle = 0.45 \lg \lambda + 10.86,
\end{equation}
dividing the clusters into two halves with the same distribution of
satellite richness $\lambda$, which we refer to simply as the
``high-$\msbcg$'' and ``low-$\msbcg$'' subsamples for the rest of the
paper. By applying the methods described in \S\ref{sec:sb} and
\S\ref{sec:sigma}, we measure the BCG+ICL surface brightness profile
$\mubi$~(top middle panel) and stellar surface mass profile $\sigbi$~(top
right panel) for each of the two subsamples.

In the top middle panel of Figure~\ref{fig:split}, solid and dashed curves
indicate the $\mubi(R)$ profiles of the high and low-$\msbcg$ subsamples,
respectively, in the SDSS $g$~(green), $r$~(red), and $i$~(maroon) bands.
By construction, the $\mubi(R)$ profile of the high-$\msbcg$ subsample is
${\sim}0.73$ magnitudes~(in the $r$ and $i$ bands; ${\simeq}0.65$
magnitudes in the $g$ band) brighter than that of the low-$\msbcg$ one on
scales well below the effective cModel aperture, i.e., $R_*{=}50\,\kpc$,
because the average $\msbcg$ of the two subsamples differ by
${\sim}0.34$ dex. However, the small-scale discrepancy between the two
subsamples persist on scales much larger than $R_*$, and the two sets of
$\mubi$ profiles do not converge within the uncertainties until
$R{\sim}300\,\kpc$. Likewise, the two $\sigbi$ profiles in the top right
panel exhibit a significant discrepancy on scales as large as
$R{\sim}200{-}300\,\kpc$, beyond which they start to converge to the same
level of stellar surface mass density.

The comparison between the two $\sigbi$ profiles is better illustrated in
the bottom of the top right panel of Figure~\ref{fig:split}, where we show
the $\sigbi$ ratios of the high~(red solid) and low~(blue dashed) $\msbcg$
subsamples over the overall sample.  The discrepancy between the two ratio
profiles is more than a factor of two across all scales below
$R{=}100\,\kpc$ and declines slowly on larger scales, but still
strongly persists at ${>}50\%$ level at $R{\sim}200{-}300\,\kpc$. Not until
distances exceed $R{\sim}300\,\kpc$ do the two ratio profiles begin to
converge to unity~(albeit with large errorbars).

Such discrepancy observed between the two subsamples provides a clear
detection of the BCG sphere of influence. Since the two subsamples of
clusters have the same richness and differ solely in their BCG stellar
mass, the clear transition from the constant discrepancy below
$R{=}100\,\kpc$ to the apparent convergence above $300\,\kpc$ in the top
right panel of Figure~\ref{fig:split} demonstrates that the BCG sphere of
influence extends to scales around $\rsoi{\simeq}200\,\kpc$.
Interestingly, such transition at $\rsoi$ coincides with the radial extent
of the transitional stellar mass component revealed by the $\sigbi$
decomposition in \S\ref{subsec:decomposition}, i.e.,
$\rsoi{\simeq}R_t{\times}10^{\sigma_t}$, suggesting a common origin of the
two observed ``transitions''. In particular, the excess diffuse light on
scales between $R_*$ and $\rsoi$ could be primarily formed via processes
that simultaneously enriched the BCG, likely due to the tidal
disruption/stripping of satellites during periapsis and the violent
relaxation after major mergers between the BCGs and satellites.
Furthermore, the convergence of the two $\sigbi$ profiles on scales above
$300\,\kpc$ confirms our expectation that the ICL in the outer region of
clusters largely follows the distribution of satellite galaxies, hence that
of the dark matter.

The bottom panels of Figure~\ref{fig:split} shows the results of a similar
experiment as in the top panels, but by dividing the clusters into low and
high-$\lambda$ subsamples by $\lambda$ at fixed $\msbcg$~(bottom left
panel). As expected, the two sets of $\mubi$ and $\sigbi$ profiles are
consistent on scales below $R_*{=}50\,\kpc$, the aperture of our $\msbcg$
measurements. The two subsamples start to differ on scales
above $R_*$, exhibiting a $20\%$ discrepancy on scales below $100\,\kpc$,
much weaker than the factor of two discrepancy between the two $\msbcg$
splits.  The two $\sigbi$ profiles do not reach a discrepancy comparable to
that exhibited by the $\msbcg$ splits until
$200\,\kpc$ at $50\%$.  This suggests that the diffuse light at
$200\,\kpc$ has equal contributions from the BCG vs. satellites,
further corroborating our conclusion that the BCG sphere of influence ends
at $\rsoi{\simeq}200\,\kpc$.  On scales between $R_*$ and $\rsoi$, the
discrepancy between the low and high-$\lambda$ clusters is significantly
smaller than that between the two $\msbcg$-split halves, indicating that
the influence of satellite richness is subdominant compared to the BCG on
those transitional scales --- they are firmly within the sphere of
influence of the BCG.

Figure~\ref{fig:image2Dsub} provides a more visually-appealing way of
comparing the diffuse light distributions between the low and high-$\msbcg$
subsamples. In particular, we compare the 2D stacked $r$-band images of the
high-$\msbcg$~(left) and low-$\msbcg$~(middle) subsamples, with the
difference image between the two shown in the right panel. The SB values,
in the unit of nanomaggies per arcsec$^{2}$, are computed with a smoothing
kernel of 15 pixels, indicated by the colourbar on the right. Solid circles
in the left and middle panels indicate the characteristic radii $r_s$ of
the high~($r_{s}{=}227.8\,\kpc$) and low~($r_{s}{=}269.7\,\kpc$) $\msbcg$
clusters, respectively, inferred from weak lensing modelling by
\citetalias{Zu2021}.  Dashed circle in each panel indicates the BCG sphere
of influence with $\rsoi{=}200\,\kpc$.  Consistent with 1D profiles in the
top panels of Figure~\ref{fig:split}, the high-$\msbcg$ clusters exhibits a
more enhanced surface brightness distribution than the low-$\msbcg$ systems
on scales below ${\sim}\rsoi$, but the two images are almost
indistinguishable on scales above ${\sim}300\,\kpc$ --- the difference
image is consistent with having zero SB at $R{>}300\,\kpc$.

\subsection{Systematic Tests Against Stellar Mass Aperture and BCG
Centering Probability}
\label{subsec:test}

\begin{figure}
    \hspace{-0.5cm}\includegraphics[width=0.48\textwidth]{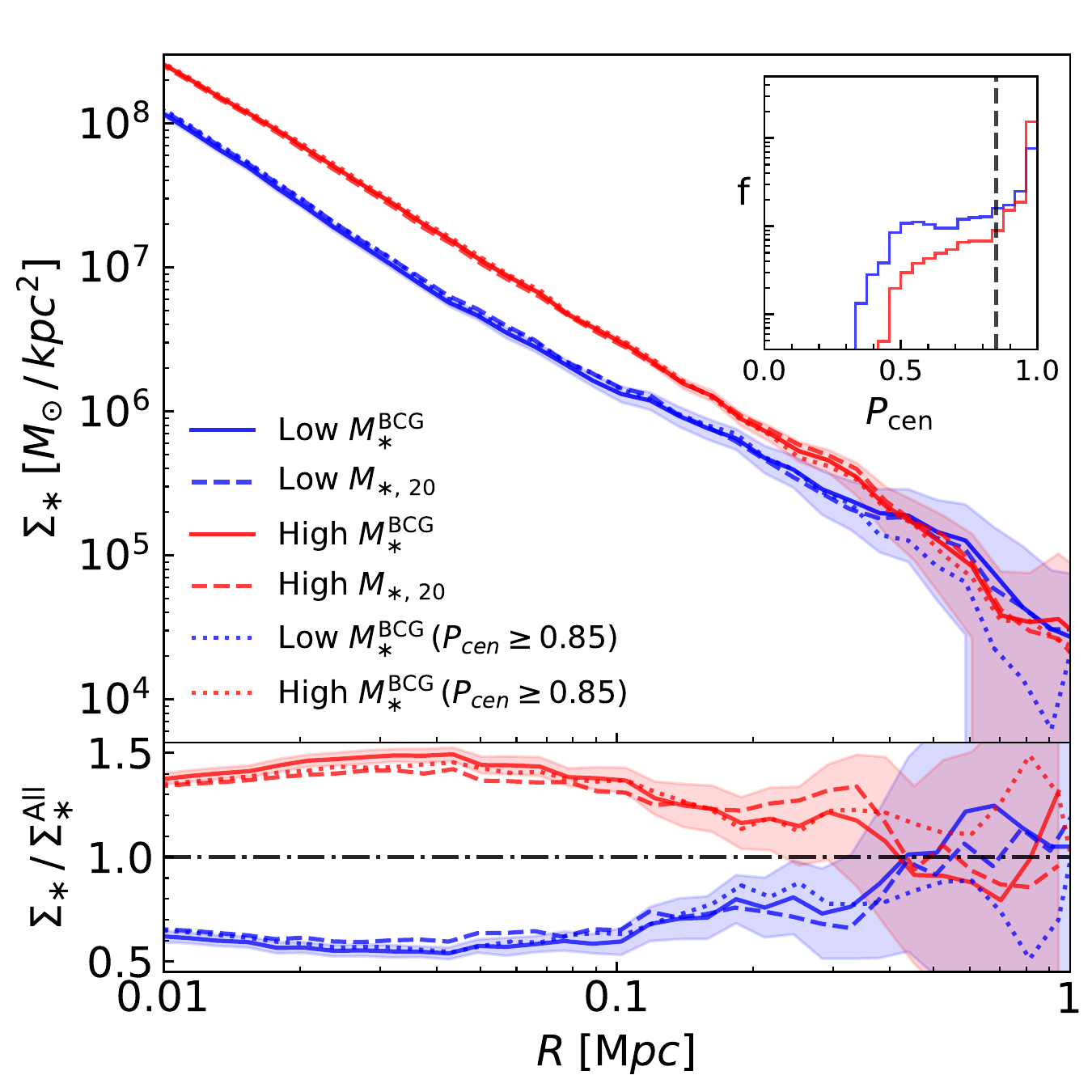}
    \caption{Similar to the top right panel of Figure~\ref{fig:split}, but
    with measurements for cluster subsamples split by $\mtwenty$~(stellar
    mass measured within a $20\,\kpc$-aperture; dashed curves) and for
    clusters with the BCG centring probability $\pcen{>}0.85$~(dotted
    curves). Inset panel shows the $\pcen$ distributions of the low~(blue
    histograms) and high~(red histograms) $\msbcg$ subsamples, with the
    vertical dashed line indicating our $\pcen{=}0.85$ cut.
    \label{fig:systest} }
\end{figure}

The key conclusion of our paper, that the BCG sphere of influence extends
to a characteristic radius of $\rsoi{\simeq}200\,\kpc$, is based on the
experiment of \S\ref{subsec:split}. However, the experiment could be
affected by systematic uncertainties associated with the aperture of our
stellar mass estimates or the miscentring effect in the \redmapper~cluster
finding algorithm~\citep{Johnston2007, Oguri2011, Rozo2014, Hollowood2019}.
For example, although the effective aperture of $\msbcg$ is
$R_*{\simeq}50\,\kpc$, the individual apertures of some of the nearby,
bright systems could be larger, thereby artificially pushing the
discrepancy between the $\sigbi$ of high and low-$\msbcg$ subsample to
larger radii.  For miscentring, the average centring probability $\pcen$ of
the low-$\msbcg$ clusters is lower~\citep[e.g., due to projection
effects; see][]{Zu2017}, and is thus more likely to have satellite galaxies
misidentified as centrals than their high-$\msbcg$ counterparts.
Consequently, it is plausible that the $\sigbi$ of the low-$\msbcg$
subsample is heavily underestimated on small scales due to the lack of
extended stellar envelope surrounding those misidentified centrals.

To investigate the impact of different stellar mass apertures, we repeat
the experiment of \S\ref{subsec:split} by adopting $\mtwenty$, the stellar
mass measured within a fixed aperture of $20\,\kpc$. By splitting the
clusters into two subsamples of different $\mtwenty$ at fixed $\lambda$, we
eliminate the possibility that the high-$\msbcg$ subsample may
preferentially select the more extended BCGs. The result of this test is
shown as the red~(high-$\mtwenty$) and blue~(low-$\mtwenty$) dashed curves
in Figure~\ref{fig:systest}.  The
$\sigbi$ of the high-$\mtwenty$ subsample is ${\sim}3\%$ lower than that
of the high-$\msbcg$ clusters on scales below $200\,\kpc$,
and vice versa for the low-$\mtwenty$ subsample.
Overall, the discrepancy between the
$\sigbi$ profiles of the two subsamples split by $\mtwenty$ is very similar
to our fiducial measurement split by $\msbcg$~(solid curves), consistent
with the BCG sphere of influence extending to $\rsoi{\simeq}200\,\kpc$.

To test the miscentring effect, we measure the $\pcen$ distributions of the
BCGs of the high~(red histograms) and low~(blue histograms) $\msbcg$
subsamples in the inset panel of Figure~\ref{fig:systest}.  Both $\pcen$
distributions peak at $\pcen$ close to 100\%, but the distribution of the
low-$\msbcg$ clusters indeed exhibits a longer low-$\pcen$ tail.  To
eliminate the impact of low-$\pcen$ systems on our $\rsoi$ measurement, we
remove clusters with the $\pcen$ of their BCGs lower than 85\%, shown as
the vertical dashed line in the inset panel, and repeat the experiment
using the $\msbcg$-split subsamples.  The results of the miscentring test
are shown by the red and blue dotted curves for the high and low-$\msbcg$
clusters with BCG $\pcen{>}85\%$, respectively. Again, we do not find any
significant deviation from our fiducial measurements due to the exclusion
of the low-$\pcen$ systems.  Therefore, based on the two systematic tests
shown in Figure~\ref{fig:systest}, we conclude that the detection of
$\rsoi{\simeq}200\,\kpc$ is robust against the aperture size of the BCG
stellar mass estimates and the level of miscentring in the SDSS
\redmapper~catalogue.

\subsection{Connection between BCG Sphere of Influence and Halo Concentration}
\label{subsec:physics}

As mentioned in the Introduction, \citetalias{Zu2021} modelled the weak
lensing measurements of the two subsamples split by $\msbcg$ at fixed
$\lambda$, and found that the two have the same average halo mass, but the
average halo concentration of the high-$\msbcg$ clusters is
${\sim}10\%$ higher than that of the low-$\msbcg$ systems~(for a thorough
explanation of such observation, see ~\citet{Zu2022}).
\citetalias{Zu2021} speculated that the strong correlation between halo
concentration and $\msbcg$ could be induced at the early phase of halo
growth, when the fast accretion and frequent mergers not only built the
characteristic central regions of the cluster haloes, but also fuel the {\it in situ}
stellar growth of the BCGs. Interestingly, the characteristic radii $r_s$
inferred by \citetalias{Zu2021} are ${\sim}200{-}300\,\kpc$, in good
agreement with our constraint of $\rsoi$ in this paper.

\begin{figure*}
    \centering\includegraphics[width=0.96\textwidth]{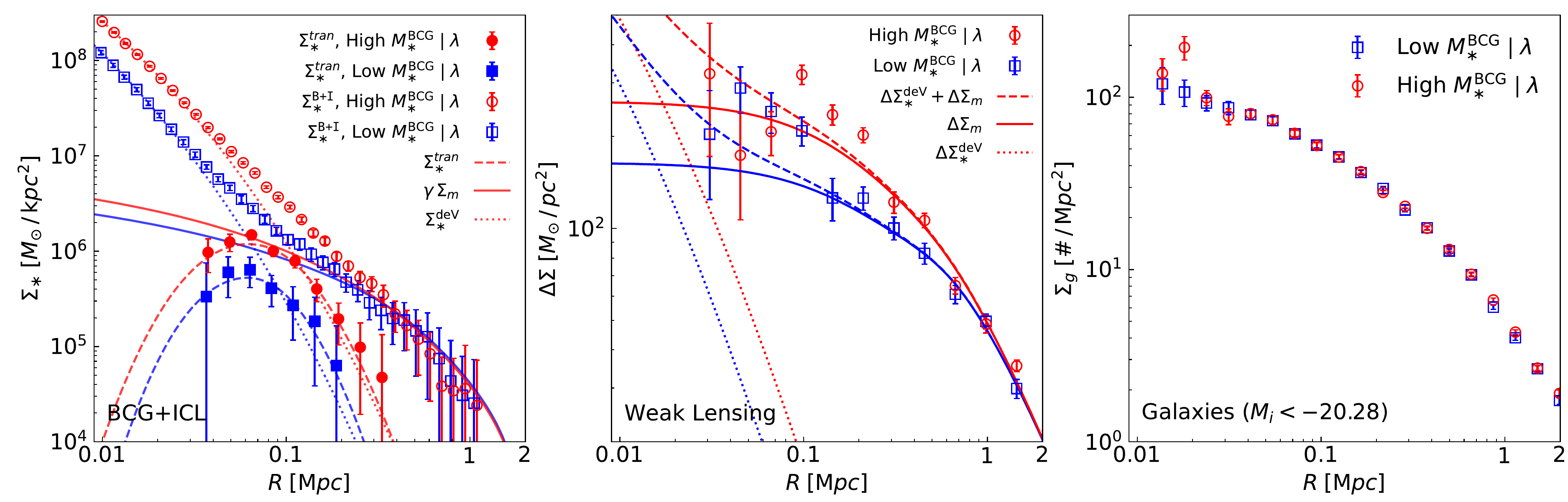}
    \caption{{\it Left}: BCG+ICL stellar surface density profiles $\sigbi$
    of the low~(blue open squares) and high~(red open circles) $\msbcg$
    subsamples. Solid and dotted curves are the best-fitting scaled matter
    density and de Vaucouleurs' profiles on relevant scales, respectively.
    Blue filled squares and red filled circles indicate the transitional
    components of the low and high-$\msbcg$ clusters, respectively, while
    the dashed curves are predictions from the best-fitting
    Equation~\ref{eqn:sigtr}.  {\it Middle}: Weak lensing profiles $\ds$ of
    the low~(blue open squares) and high (red open circles) $\msbcg$
    subsamples. Solid curves are the predictions by the best-fitting $\ds$
    models described in \citetalias{Zu2021}. The two $\ds$ models have the
    same halo mass, but the average halo concentration of the high-$\msbcg$
    subsample is ${\sim}10\%$ higher than that of the low-$\msbcg$ clusters. Dotted
    curves indicate the predicted $\ds$ signal induced by the respective de
    Vaucouleurs' profiles shown in the left panel, and each dashed curve is
    the sum of the dotted and solid curves of the same colour.  {\it
    Right}: The galaxy surface number density profiles $\sigg$ of the
    low~(blue open squares) and high (red open circles) $\msbcg$
    subsamples. The two measured profiles are consistent with each other on
    all scales.  \label{fig:subdecomposition}}
\end{figure*}

Such similarity between the values of $r_s$ and $\rsoi$ is likely physical.
In the vicinity of the BCGs, stars could be tidally disrupted/stripped from
satellites after periapsis passage or ejected after BCG-satellite major
mergers. Most of those stray stars cannot escape to larger radii due to the
relatively high escape velocities within $r_s$. Therefore, a fraction of
them would fall back onto the BCGs and grow $\msbcg$, while the rest stays
unbound to the BCGs and form the transitional component $\sigtr$, which is
however well confined within $r_s$, hence the BCG sphere of influence
$\rsoi$.  In this scenario, the BCG sphere of influence is primarily shaped
by the growth of the BCG, but helped maintained by the concentration of the
dark matter halo. Therefore, we expect that $\rsoi$ increases with the
growth of $\msbcg$, but should always stay within the characteristic radius
of the halo $r_s$.

We present a comprehensive comparison between the low~(blue squares) and
high~(red circles) $\msbcg$ subsamples in their BCG+ICL stellar surface
density profiles $\sigbi$~(left), weak lensing profiles $\ds$~(middle), and
galaxy surface number density profile $\sigg$~(right) in
Figure~\ref{fig:subdecomposition}. The left panel is similar to
Figure~\ref{fig:decomposition}, but with our decomposition method applied
to the two subsamples separately. Since we adopt the same $\gamma{=}1/446$
as for the overall sample, the discrepancy in the total surface matter
density profiles $\gamma\sigm$ between the two subsamples below $200\,\kpc$
is entirely due to the different halo concentrations.  After subtracting
the inner de Vaucouleurs' profile and the outer scaled matter surface
density profile from each $\mubi$ profile, the transitional component
$\sigtr(R)$ emerges as the filled symbols with errorbars that we fit using
Equation~\ref{eqn:sigtr}~(dashed curves).

Focusing on the $\sigtr(R)$ profiles in the left panel of
Figure~\ref{fig:decomposition}, we detect a shift of $26\,\kpc$ in the
centroid from the low-$\msbcg$ subsample~($R_{t}{=}92\,\kpc$) to the
high-$\msbcg$ subsample~($R_{t}{=}118\,\kpc$), while the two log-widths are
comparable~($\sigma_t{=}0.22$ and $0.25$ dex for low and high-$\msbcg$
subsamples, respectively). Assuming
$\rsoi{\simeq}R_{t}{\times}10^{\sigma_t}$, $\rsoi$ increases with $\msbcg$
from $R_{\mathrm{SOI}}^{\mathrm{low}}{=}153\,\kpc$ to
$R_{\mathrm{SOI}}^{\mathrm{high}}{=}210\,\kpc$. Moreover, the integrated
mass within the transitional component $\sigtr$ of the high-$\msbcg$
subsample is $8.56{\times}10^{10}\msol$~($f_t{=}0.29$), about 0.48 dex higher
than that of the low-$\msbcg$ subsample with
$2.81{\times}10^{10}\msol$~($f_t{=}0.24$) --- slightly larger than the 0.34
dex difference between average $\msbcg$.

The middle panel of Figure~\ref{fig:subdecomposition} is similar to the
Figure 5 of \citetalias{Zu2021}, showing the weak lensing comparison
between the two subsamples. We include the weak lensing signals
predicted on small scales by the de Vaucouleurs' profiles of the
BCGs~(dotted curves) in the prediction of the total $\ds$~(dashed curves).
The $10\%$ difference in halo concentration is
manifested by the discrepancy between the two $\ds$ profiles on scales
between $100\,\kpc$ and $500\,\kpc$, which then translates to the
discrepancy in their total surface matter density profiles $\sigm$ below
$200\,\kpc$ in the left panel. We compare the two galaxy surface number
density profiles in the right panels of
Figure~\ref{fig:subdecomposition}~(similar to the right panel of Figure 6
in \citetalias{Zu2021}). Clearly, the galaxy distributions around the BCGs
of the two subsamples are almost indistinguishable, further corroborating
our conclusion that the differences in $\sigbi$~(left) and $\ds$~(middle)
are tied solely to the discrepancy in $\msbcg$ and/or halo concentration.

\section{Conclusion}
\label{sec:conc}

In this paper, we have measured the stacked BCG+ICL surface brightness
profiles $\mubi(R)$ in the SDSS $gri$ bands for a sample of ${\sim}3000$
clusters detected between $0.2{<}z{<}0.3$ from SDSS DR8 imaging. Adopting
an empirically calibrated mass-to-light relation, we convert the three-band
$\mubi$ profiles to the diffuse stellar surface density profile
$\sigbi(R)$. By comparing the $\sigbi$ profile with the cluster weak
lensing profile, we find that the ICL on scales $R{>}400\,\kpc$ can be well
described by the projected dark matter density profile $\sigm$ multiplied
by the ICL-to-halo mass ratio of $1/675$.

Further assuming that the inner BCG follows the de Vaucouleurs' law, we
develop a physically-motivated method to decompose $\sigbi(R)$ into three
components, including an inner de Vaucouleurs' profile, an outer ICL that
follows the dark matter, and a third excess component at the transitional
scales between $70\,\kpc$ and $200\,\kpc$. We find that the ratio between
this transitional component $\sigtr$ and $\sigbi(R)$ can be well described
by a Gaussian function that peaks at $R_t{=}116\,\kpc$ with a logarithmic
dispersion of $\sigma_t{=}0.23$ dex.  After correcting for mask
incompleteness using the conditional stellar mass function of satellites
measured by \citet{Yang2012}, we infer the stellar mass
budget of the clusters in three states as: the BCG~($8.4\%$; de
Vaucouleurs + transitional), the satellites~($83.5\%$), and the diffuse ICL
that follows the dark matter distribution~($8.1\%$). By measuring the
weak lensing halo mass as $2.57{\times}10^{14}\,\msol$, the stellar-to-halo
mass fractions are $1.531\%$~(satellites), $0.155\%$~(BCG), and
$0.148\%$~(ICL), leading to a total stellar-to-halo fraction of $1.834\%$.

The transitional component $\sigtr$ is the key to solving the radius of the
BCG sphere of influence $\rsoi$ in the diffuse cluster light --- $\rsoi$
could be well within $R_t{\times}10^{-\sigma_t}{\sim}70\,\kpc$ if $\sigtr$
is induced by the interaction between satellite galaxies and the cluster
potential, or extend to $R_t{\times}10^{\sigma_t}{\sim}200\,\kpc$ if
$\sigtr$ is tied to the BCG growth. To distinguish the two physical
scenarios, we divide the clusters into two subsamples by their BCG stellar
mass $\msbcg$~(with an effective aperture of ${\sim}50\,\kpc$) at fixed
satellite richness $\lambda$, and compare their BCG+ICL surface brightness
and stellar density profiles.  Surprisingly, we discover that the two
$\sigbi$ profiles differ significantly on all scales below
$R{\sim}200\,\kpc$, before converging to the same amplitude on larger
scales.

From the weak lensing and cluster-galaxy cross-correlations in
\citetalias{Zu2021}, we show that the two subsamples have the same average
halo mass as well as the same satellite number
density profile, but differ only in their average BCG stellar mass.
Therefore, the observed discrepancy between the two $\sigbi$ profiles
signals a detection of the BCG sphere of influence at
$\rsoi{\simeq}200\,\kpc$.  We test the sensitivity of our experiment to the
aperture size of stellar mass measurements as well as the miscentring
effect of the optical clusters, and demonstrate that $\rsoi{\simeq}200\,\kpc$ is
robust against the two systematic uncertainties in the experiment.

Weak lensing measurements also suggest that the high-$\msbcg$ clusters are
${\sim}10\%$ more concentrated than their low-$\msbcg$ counterparts,
suggesting that $\rsoi$ is likely connected to the characteristic scale
$r_s$ of the NFW haloes. We speculate that the relatively high escape
velocity profile within $r_s$ may help confine the stray stars that break
free of the BCGs, producing the excess mass at the transitional
scales~\citep{Kelson2002, Bender2015, Veale2018}. Detailed kinematic analysis of the hydrodynamical simulations of cluster
formation could be tremendously helpful for providing insight to the growth
of the BCG sphere of influence, as well as an indirect check of the
physical link between $r_s$ and $\rsoi$. Observationally, our work can be
easily extended to the larger cluster catalogues~\citep{Zou2021, Wen2021,
Yang2021} and deeper photometry~\citep{Huang2021, Li2021} from current
stage-III imaging surveys.  Future ground-based and space missions like the
LSST, {\it Euclid}, {\it Roman}, and {\it CSST} with even larger cluster
samples, deeper photometry, and sharper shear measurements will greatly
enhance our capability of seeing cluster formation through the diffuse
light.

\section*{Data availability}

The data underlying this article will be shared on reasonable request to the corresponding author.

\section*{Acknowledgements}

We thank the anonymous referee for suggestions that have substantially
improved the manuscript, and Weiguang Cui, Fengshan Liu, Zhonglue Wen, and
Hu Zou for helpful discussions. Y.Z. acknowledges the support by the
National Key Basic Research and Development Program of China (No.
2018YFA0404504), National Science Foundation of China (11873038, 11621303,
11890692, 12173024), and the science research grants from the China Manned
Space Project (No.  CMS-CSST-2021-A01, CMS-CSST-2021-A02,
CMS-CSST-2021-B01). Y.Z. acknowledges the National One-Thousand Youth
Talent Program of China, the SJTU start-up fund (No.  WF220407220), and the
support by the 111 Project of the Ministry of Education under grant No.
B20019. H.S. acknowledges the support from CMS-CSST-2021-A01, NSFC of China
under grant 11973070, the Shanghai Committee of Science and Technology
grant No.19ZR1466600, and Key Research Program of Frontier Sciences, CAS,
Grant No. ZDBS-LY-7013.  X.C.  acknowledges the hospitality of Zhejiang
University where he enjoyed a fruitful discussion with participants of the
23rd ``Guoshoujing Meeting on Galaxy Formation and Cosmology''.

%%%%%%%%%%%%%%%%%%%% REFERENCES %%%%%%%%%%%%%%%%%%

% The best way to enter references is to use BibTeX:

% \clearpage
\bibliographystyle{mnras}
% \interlinepenalty=10000
%\bibliography{ref} %
% \bibliography{ref_me.bib}

% Alternatively you could enter them by hand, like this:
% This method is tedious and prone to error if you have lots of references
% \begin{thebibliography}{99}
% \bibitem[\protect\citeauthoryear{Author}{2012}]{Author2012}
% Author A.~N., 2013, Journal of Improbable Astronomy, 1, 1
% \bibitem[\protect\citeauthoryear{Others}{2013}]{Others2013}
% Others S., 2012, Journal of Interesting Stuff, 17, 198
% \end{thebibliography}

%%%%%%%%%%%%%%%%%%%%%%%%%%%%%%%%%%%%%%%%%%%%%%%%%%

% Don't change these lines
\bsp	% typesetting comment
\label{lastpage}

\end{document}